\documentclass[pra,10pt,nofootinbib, onecolumn,notitlepage, superscriptaddress]{revtex4-1}

\pdfoutput=1

\usepackage{amsmath}
\usepackage{amsfonts}
\usepackage{braket}
\usepackage{amssymb}
\usepackage{mathbbol}
\usepackage{graphicx}

\usepackage[colorlinks=true,linkcolor=blue, citecolor=blue, urlcolor=blue, bookmarks]{hyperref}
\def\be{\begin{equation}}
	\def\ee{\end{equation}}
\def\bea{\begin{eqnarray}}
	\def\eea{\end{eqnarray}}

\begin{document}
	
	\title{A Random Unitary Circuit Model for Black Hole Evaporation}
	
	\author{Lorenzo Piroli$^\ast$}
	\affiliation{Max-Planck-Institut f\"ur Quantenoptik, Hans-Kopfermann-Str.~1, 85748 Garching, Germany}
	\affiliation{Munich Center for Quantum Science and Technology, Schellingstraße 4, 80799 M\"unchen, Germany}
	\author{Christoph S\"underhauf$^\ast$}
	\affiliation{Max-Planck-Institut f\"ur Quantenoptik, Hans-Kopfermann-Str.~1, 85748 Garching, Germany}
	\affiliation{Munich Center for Quantum Science and Technology, Schellingstraße 4, 80799 M\"unchen, Germany}
	\author{Xiao-Liang Qi}
	\affiliation{Stanford Institute for Theoretical Physics, Stanford University, Stanford, California 94305, USA}
	\affiliation{Department of Physics, Stanford University, Stanford, California 94305, USA}
	
	\begin{abstract}
		
	Inspired by the Hayden-Preskill protocol for black hole evaporation, we consider the dynamics of a quantum many-body qudit system coupled to an external environment, where the time evolution is driven by the continuous limit of certain $2$-local random unitary circuits. We study both cases where the unitaries are chosen with and without a conserved $U(1)$ charge and focus on two aspects of the dynamics. First, we study analytically and numerically the growth of the entanglement entropy of the system, showing that two different time scales appear: one is intrinsic to the internal dynamics (the scrambling time), while the other depends on the system-environment coupling. In the presence of a $U(1)$ conserved charge, we show that the entanglement follows a Page-like behavior in time: it begins to decrease in the middle stage of the ``evaporation'', and decreases monotonically afterwards. Second, we study the time needed to retrieve information initially injected in the system from measurements on the environment qudits. Based on explicit numerical computations, we characterize such time both when the retriever has control over the initial configuration or not, showing that different scales appear in the two cases.
	\end{abstract}
	
	\maketitle
	
	\def\thefootnote{*}\footnotetext{These authors contributed equally to this work.}\def\thefootnote{\arabic{footnote}}
		
	%%%%%%%%%%%%%%%%%%%%%%%%%%%%%%%%%%
	
	\section{Introduction}
	\label{sec:intro}
	
	In the past decade, quantum information ideas have become increasingly relevant in high energy physics, especially in connection to the black hole information paradox~\cite{Hawking1975,Hawking1976,almheiri2013black,penington2019entanglement,almheiri2019entropy}. In this context, a particularly fruitful line of research was initiated by the seminal work by Hayden and Preskill~\cite{hayden2007black}, where the authors studied how quantum information is released from a black hole, under the assumption that it is not destroyed during the evaporation process. Their study suggested that information could be released in a time which is much shorter than the black hole lifetime, and related to the time needed for localized information to spread, or \emph{scramble}, over all the degrees of freedom. 
	
	These considerations provided an obvious motivation for a systematic study of information scrambling and the related concept of many-body quantum chaos, also due to the subsequent conjecture by Sekino and Susskind that black holes are the fastest scramblers in nature~\cite{sekino2008fast,susskind2011addendum}. In turn, this led to the development of several measures of information spreading and chaos, including out-of-time-ordered correlation (OTOC) functions~\cite{shenker_black_2014, shenker_multiple_2014,Kitaev_talk, roberts_localized_2015,maldacena2016bound,roberts_operator_2018} (historically introduced in the context of disordered superconductors~\cite{larkin1969quasiclassical}), and the tripartite mutual information defined in Ref.~\cite{hosur_chaos_2016}.
	
	Due to the intrinsic complexity of generic many-body quantum systems, several works on the topic relied on the study of  a class of simplified dynamical models given by random unitary circuits (RUCs), originally introduced within quantum information theory~\cite{emerson_convergence_2005,dahlsten_emergence_2007,gross_evenly_2007,znidaric_optimal_2007,znidaric_exact_2008,arnaud_efficiency_2008,harrow_random_2009,brown_convergence_2010,diniz_comment_2011,brandao_local_2016,nakata_efficient_2017}, and continuous Brownian  dynamics~\cite{lashkari2013towards,onorati_mixing_2017,banchi_driven_2017} . These models are generally defined in terms of a set of $d$-level systems (qudits), sequentially updated by randomly chosen unitary gates (RUCs) or time-dependent random Hamiltonians (continuous Brownian dynamics). It turns out that these systems are typically fast scramblers~\cite{lashkari2013towards}, and their study allowed us to investigate quantitatively several interesting features that are expected in more realistic chaotic systems~\cite{lashkari2013towards, gharibyan_onset_2018,saad2018semiclassical,zhou_operator_2019,sunderhauf_quantum_2019}. As a parallel development, these ideas also had important ramifications in condensed matter and many-body physics, where local RUCs have been extensively studied in the past few years~\cite{nahum2017quantum,sunderhauf2018localization,nahum2018operator,vonKeyserlingk2018operator,rakovszky2018diffusive,chan2018solution,chan2018spectral,khemani2018operator,Kos2018Many,Bertini_Exact2018,hunter2019unitary,Gullans2017Entanglement,zhuang2019_Scrambling}, for instance in connection with  aspects of entanglement spreading and thermalization in isolated systems~\cite{d2016quantum,rigol2008thermalization}. 
	
	In this work, motivated by the recent technical advances in the study of RUCs, and inspired by the Hayden-Preskill evaporation protocol, we consider the dynamics of a quantum many-body qudit system coupled to an external environment, where the time evolution is driven by the continuous limit of certain $2$-local RUCs. These consist of qudits nonlocally coupled, but with only two of them interacting at a time. This setting allows us to study quantitatively the contribution of the environment and internal dynamics on the scrambling of information. Furthermore, we consider a modified tensor network model with $U(1)$ charge conservation, which evaporates to a unique vacuum state, instead of reaching the maximally entangled state. This provides a more realistic toy model of evaporating black hole in flat space, for which the entropy after the Page time eventually decreases to zero~\cite{PageInformation_1993,page2005hawking}. The $U(1)$ charge conservation is an analog of the energy conservation. 
	
	In the rest of this paper, we focus on two aspects of the dynamics. First, we study analytically and numerically the growth of the second R\'enyi entropy of the system, highlighting the implications of conservation laws and the emergence of two different time scales: one is intrinsic to the internal dynamics (the scrambling time), while the other depends on the system-environment coupling. Second, following Hayden and Preskill~\cite{hayden2007black}, we study the time needed to retrieve information initially injected in the system from measurements on the environment qudits, and how this depends on the knowledge of the initial configuration of the system.
	
	In the past years, several works have appeared discussing ideas and techniques related to those of the present  paper. First, we note that our setting differs from those studied in Refs.~\cite{li_quantum_2018,li_measurement-driven_2019,skinner_measurement-induced_2019,
		gullans_dynamical_2019,zabalo_critical_2019,bao_theory_2019,choi_quantum_2019,
		vasseur_entanglement_2019,gullans_scalable_2019} in the context of  measurement-induced phase transitions.  Indeed, in our model no projective measurement is taken, and we consider instead an environment which is eventually traced over in our calculations. A similar setting was studied in Ref.~\cite{chernowitz2020entanglement}, but there the authors considered random global Hamiltonians, with no notion of local interactions. Next, quantum mechanical evaporation protocols displaying some analogy with our setting were investigated in Refs.~\cite{almheiri2019universal,ZhangEvaporation2019} for an SYK model~\cite{Sachdev1993Gapless,Kitaev_talk} coupled to an external environment (see also \cite{chen2017tunable}). However, the dynamics studied in these works is not Brownian, and is analyzed by means of the Keldysh formalism. 
	
	It is also worth to stress that over the years many qudit models have been introduced to capture aspects of the black hole evaporation process~\cite{mathur_information_2009,czech_black_2011,mathur2011correlations,mathur_information_2011,cai_comment_2012,brady_scrambling_2013,giddings_models_2012,avery_unitarity_2013,giddings_quantum_2013,avery_qubit_2013,verlinde2013passing,roy_does_2014,hubener_equilibration_2015,bradler_one-shot_2016,leutheusser_tensor_2017,tokusumi_quantum_2018,alvi_modifications_2019}. In particular, a Page-like behavior in time for the entanglement entropy has already been observed in some of these~\cite{czech_black_2011,mathur2011correlations,bradler_one-shot_2016,tokusumi_quantum_2018}. However, in most of these examples the evolution is  very carefully engineered and allows one to only study numerically small system sizes. 
	
	We also mention that very recently the effects of decoherence on information scrambling has been analyzed in Ref.~\cite{Yoshida2019Disentangling} within a quantum teleportation protocol related to the setting of this paper, see also Ref.~\cite{landsman2019verified} for an experimental implementation. Furthermore, we note that the Hayden-Preskill protocol with a $U(1)$ conserved charge has been studied before in Ref.~\cite{Yoshida2019Soft}, where global random unitary transformations (instead of $k$-local circuits) were considered. Finally, two papers closely related to the present article appeared very recently. First,  a random quantum circuit model for black hole evaporation was studied in Ref.~\cite{agarwal_toy_2019}, but there the authors focused on a different setup and quantities . Second, analogously to our work, the emergence of a Page curve in a unitary toy model for a black hole has also been shown in Ref.~\cite{liu_dynamical_2020}, based on recently-developed concepts of many-body quantum chaos. However, in this work we focus on a specific microscopic model which is different from the one studied in Ref.~\cite{liu_dynamical_2020}, and employ different techniques in our calculations.
	
	The rest of this manuscript is organized as follows. In Sec.~\ref{sec:model} we introduce our model, while in Sec.~\ref{sec:entanglement_growht} we analyze the growth of the entanglement both in the case of Haar-scrambled local unitary evolution (Sec.~\ref{sec:entropy_no_cons}) and in the presence of a $U(1)$ conserved charge (Sec.~\ref{sec:entropy_u1}). The retrieval of quantum information initially injected in the system is studied in Sec.~\ref{sec:info_retrieval}, while we report our conclusions in Sec.~\ref{sec:conclusions}. Finally, the most technical aspects of our work are consigned to a few appendices.

	\section{The model}
	\label{sec:model}
	
	\begin{figure}
		\includegraphics[width=0.4\linewidth,height=0.3\linewidth]{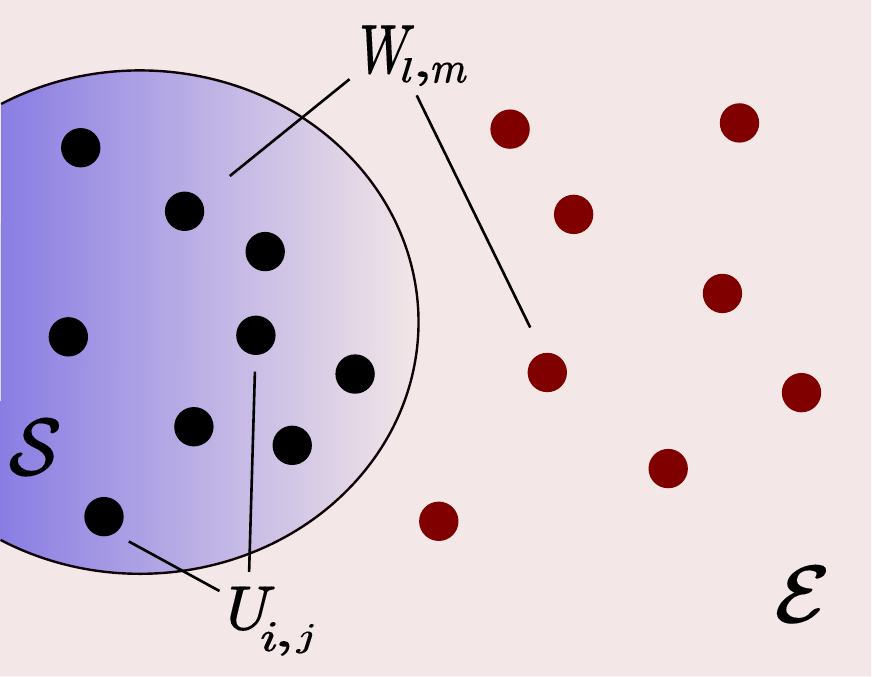}
		\caption{Pictorial representation of the model introduced in  Sec.~\ref{sec:model}. We consider a system $\mathcal{S}$ and an environment $\mathcal{E}$ consisting of $N$ and $M$ qudits respectively. The evolution  is driven by the continuous limit  of a random quantum circuit  which implements a fast-scrambling dynamics for $\mathcal{S}$ with a tunable coupling between $\mathcal{S}$ and $\mathcal{E}$. At each infinitesimal time step $\Delta t$ a random unitary operator $U_{i,j}$ is applied to randomly chosen qudits in $\mathcal{S}$ with probability $p_1=N\lambda_1\Delta t$, while a swap $W_{l,m}$ between a qudit in the system and one in the environment (randomly chosen) is applied  with probability  $p_2=\lambda_2\Delta  t$.}
		\label{fig:model}
	\end{figure}
	
	We start by introducing the model studied in the rest of this work, which is pictorially depicted in Fig.~\ref{fig:model}. We consider two sets of $N$ and $M$ $d$-level systems (qudits), denoted respectively by $\mathcal{S}$ (the system) and  $\mathcal{E}$ (the environment). The Hilbert spaces associated with $\mathcal{S}$ and $\mathcal{E}$ are then $\mathcal{H}_{\mathcal{S}}=\bigotimes_{j=1}^N h^{(j)}_{\mathcal{S}}$ and $\mathcal{H}_{\mathcal{E}}=\bigotimes_{j=1}^M  h^{(j)}_{\mathcal{E}}$, with $h^{(j)}_{\mathcal{S}}, h^{(j)}_{\mathcal{E}}\simeq \mathbb{C}^d$. We anticipate that in our calculations we will always take the limit $M\to\infty$, corresponding to the physical situation where the number of degrees of freedom in the environment is much larger than in the system.
	
	Motivated by the Hayden-Preskill evaporation protocol~\cite{hayden2007black}, we would like to construct a quantum  circuit which implements  a fast-scrambling dynamics for $\mathcal{S}$  and with a tunable coupling between $\mathcal{S}$ and $\mathcal{E}$. Let us begin by considering a discrete process, and divide the time interval $[0,t]$ into $n$  steps $t_j=(j/n)t$, so that $t_j-t_{j-1}=\Delta t= t/n$. At each time step, the system evolves according to the following rules: 
	\begin{enumerate}
		\item with probability $p_1$, two qudits in $\mathcal{S}$, placed  at random positions $i$ and $j$, interact. We model this process by the action on $h^{(i)}_{\mathcal{S}}\otimes h^{(j)}_{\mathcal{S}}$ of a unitary operator $U_{i,j}$, chosen out of a suitable random ensemble;
		\item with probability $p_2\leq 1-p_{1}$, one qudit in $\mathcal{S}$ and one qubit in $\mathcal{E}$ at random positions are swapped. This models the  simplest possible interaction between $\mathcal{S}$ and $\mathcal{E}$.
	\end{enumerate}
	Note that at each time step the system is not evolved with  probability $1-p_1-p_2$. The random choice of interacting qudits should be considered as ``fixed once chosen": as we will see later, this means that when multiple replicas of the system are considered, the circuit is always identical in each copy.
	
	The  above rule defines a quantum circuit with discrete time steps. It is convenient to  take a continuous limit of the former, which  allows us to simplify some aspects  of the computations. In order to do so, we choose the probability $p_1$ and  $p_2$ to scale with the time interval $\Delta t$ as
	\bea
	p_1 &=& N\lambda_1\Delta t\, \label{eq:p_1},\\
	p_2&=& \lambda_2\Delta t\,,\label{eq:p_2}
	\eea
	where $\lambda_1$, and $\lambda_2$ are two positive real numbers. Note that while both $p_1$ and $p_2$ are proportional to $\Delta t$, they have a different dependence on $N$. As we will comment on again later, this ensures that the internal time scales are much shorter that those related to  the interaction with the environment, as it is assumed within the Hayden-Preskill protocol~\cite{hayden2007black}.  With the above choices, expectation values of observables  computed at time $t$ display a well defined limit for  $\Delta t\to 0$ (namely  $n\to \infty$), yielding a continuous dynamics for $\mathcal{S}\cup \mathcal{E}$. Importantly, we will be interested in the limit of an infinitely large environment, which will then play the role of a ``qudit'' reservoir. In the discrete dynamics, it is enough to choose the number $M$ of environment qudits to be $M\gg N t/\Delta t$, so that $M\to \infty$ in the continuous limit.
	
	In the rest of this work, we will focus on the computation of \emph{averaged} physical quantities: at each time step this amounts to averaging over all the possible choices of pairs of qudits and of gates $U_{i,j}$, with the proper probability distribution. For a  given fixed time $t$,  this is equivalent to averaging over all the realizations of allowed quantum circuits. A crucial point is that each individual realization corresponds to a unitary  evolution. In particular, if the initial state of $\mathcal{S}\cup \mathcal{E}$ is pure, it will remain so for any realization, and its von Neumann entanglement entropy will remain zero for all times (and so will its average over realizations). 
	
	Finally, regarding the ensemble of two-qudit gates $U_{i,j}$, we will consider two distinct physical situations. In the first one, the internal dynamics is ``maximally chaotic'', namely each gate $U_{i,j}$ is drawn out of a Haar distribution. In the second situation, we assume a locally conserved $U(1)$ charge, namely we choose each gate $U_{i,j}$ to preserve the $U(1)$ sectors in  the product $h^{(i)}_{\mathcal{S}}\otimes h^{(j)}_{\mathcal{S}}$, as it was done in Refs.~\cite{khemani2018operator,rakovszky2018diffusive} for the case of spatially local RUCs.

	\section{The entanglement growth}
	\label{sec:entanglement_growht}
	
	In this section we study the entanglement growth for a subsystem $K\subset\mathcal{S}$, which is naturally quantified by means of the von Neumann entanglement entropy 
	\be
	S_{K}(t)=-{\rm tr}_K\left[\rho_K(t)\log_2 \rho_K(t) \right]\,,
	\ee
	where $\rho_K(t)$ is the density matrix reduced to the subsystem $K$. Denoting by $\{\ket{j}\}_{j=0}^{d-1}$, a basis for the local Hilbert spaces $h_\mathcal{S}$,$h_\mathcal{E}$, we will assume that both the system and the environment are initialized in product states, denoted by $|\Psi^{\mathcal{S}}_{0}\rangle$ and $|\Psi^{\mathcal{E}}_{0}\rangle$ respectively. In particular we set, for finite $M$,
	\be
	\ket{\Psi^{\mathcal{E}}_0}=\ket{0}^{\otimes M}\,, \label{eq:initial_state_environment}
	\ee
	while we will consider different initial product states for $\mathcal{S}$. Note that by construction there is no entanglement between $\mathcal{S}$ and $\mathcal{E}$ at time $t=0$.
	
	Despite the importance of the  von Neumann entanglement entropy, it is known that the latter is difficult to obtain in the setting of RUCs~\cite{nahum2017quantum}. For this reason, in the following we focus on the related R\'enyi-$2$ entropy; more precisely, we will compute
	\be
	S^{(2)}_K(t)=-\log_2\left\{{\rm tr}_K\left\{\mathbb{E}\left[\rho^2_K(t)\right]\right\}\right\}\,.
	\label{eq:log_purity}
	\ee
	We note that $S^{(2)}_K(t)$ is not the averaged second R\'enyi entropy, as the
	disorder average is taken inside the logarithm. In fact, Eq.~\eqref{eq:log_purity} is proportional to the logarithm of the averaged purity $\mathcal{P}_K$, which is defined as
	\be
	\mathcal{P}_K(t)={\rm tr}_K\left\{\mathbb{E}\left[\rho^2_K(t)\right]\right\}\,.
	\ee
	However, for large $N$ one expects the effect of fluctuations in the disorder to be small, so that the behavior of $S^{(2)}_K(t)$  should be qualitatively the same as the averaged R\'enyi-$2$ entropy~\cite{vonKeyserlingk2018operator}. 
	
	Let us now define $\overline{K}=\mathcal{S}\setminus K$ and rewrite
	\bea
	{\rm tr}_K\left\{\mathbb{E}\left[\rho^2_K(t)\right]\right\}&=&{\rm tr}_K\left\{\mathbb{E}\left[{\rm tr}_{\overline{K}}\left[\rho_{\mathcal{S}}(t)\right]{\rm tr}_{\overline{K}}\left[\rho_{\mathcal{S}}(t)\right]\right]\right\}\nonumber\\
	&=&{\rm tr}_{K\otimes K}\left\{X_K\,{\rm tr}_{\overline{K}\otimes \overline{K}}\left\{ \mathbb{E}\left[\rho_{\mathcal{S}}(t)\otimes \rho_{\mathcal{S}}(t) \right]\right\}\right\}\,\nonumber\\
	&=&{\rm tr}\left\{X_K\mathbb{E}\left[\rho_{\mathcal{S}}(t)\otimes \rho_{\mathcal{S}}(t)\right]\right\}
	\eea
	where $X_K$ is a swap operator exchanging the two copies of $K$, while in the last line, ``${\rm tr}$'' represents the trace over the entire Hilbert space. From this expression it is clear that $S^{(2)}_K(t)$ is completely determined by the knowledge of $\mathbb{E}\left[\rho_{\mathcal{S}}(t)\otimes \rho_{\mathcal{S}}(t) \right]$. In order to compute the latter, it is convenient to recall the  Choi-Jamiolkowski mapping which allows us to interpret the operator $\rho_{\mathcal{S}}(t)\otimes \rho_{\mathcal{S}}(t) $ (defined on the tensor product of two ``replicas'' $\mathcal{H}_\mathcal{S}\otimes \mathcal{H}_\mathcal{S}$ ) as a state in $\mathcal{H}_{\mathcal{S}}^{\otimes 4}$. In particular, we define
	\be
	\ket{\rho_{\mathcal{S}}(t)\otimes \rho_{\mathcal{S}}(t)}\rangle=\left(\mathbb{1}_{\mathcal{H}_S}\otimes \rho_{\mathcal{S}}  (t)\otimes \mathbb{1}_{\mathcal{H}_S}\otimes \rho_{\mathcal{S}}  (t)\right)\ket{I^{+}}_{1,\ldots , N}\,,
	\label{eq:choi_map}
	\ee
	where we introduced the maximally entangled state
	\be
	\ket{I^+}_{1,\ldots , N}=\bigotimes_{k=1}^{N}\ket{I^+}_{k}\,,
	\ee
	with
	\be
	\ket{I^+}_{k}=\sum_{a,b=0}^{d-1}\left(\ket{a}_{k}\otimes \ket{a}_{k}\right)\otimes \left(\ket{b}_k\otimes \ket{b}_k\right)\,.
	\label{eq:def I+}
	\ee
	In the following, we label with $1$ and $2$ the Hilbert spaces of the two replicas associated with $\rho_{S}(t)$ in Eq.~\eqref{eq:choi_map}, and with $\bar{1}$ and $\bar{2}$ the other two. Accordingly, the Hilbert space corresponding to the four replicas is
	\be
	\widetilde{\mathcal{H}}_{\mathcal{S}}=\mathcal{H}^{(\bar{1})}_{\mathcal{S}}\otimes \mathcal{H}^{(1)}_{\mathcal{S}}\otimes \mathcal{H}^{(\bar{2})}_{\mathcal{S}}\otimes \mathcal{H}^{(2)}_{\mathcal{S}}\,.
	\ee
	Finally, we also define
	\begin{equation}
	\ket{I^-}_k = \sum_{a,b=0}^{d-1}(\ket{a}_k\otimes\ket{b}_k)\otimes(\ket{b}_k\otimes\ket{a}_k).
	\end{equation}
	Within this formalism one can recover the value of the purity using
	\be
	\mathcal{P}_K(t) \equiv {\rm tr}_K\left\{\mathbb{E}\left[\rho^2_K(t)\right]\right\}=\langle\braket{W_{K}|\mathbb{E}\left[\rho_\mathcal{S}(t)\otimes \rho_\mathcal{S}(t)\right]}\rangle\,,
	\label{eq:def_purity}
	\ee
	where 
	\be
	\ket{W_K}\rangle= \bigotimes_{k\in K}\ket{I^{-}}_k\bigotimes_{k\in \overline{K}}\ket{I^{+}}_k\,.
	\label{eq:vector_form}
	\ee
	Eq.~\eqref{eq:def_purity} can be verified straightforwardly by expanding the scalar product. We note that when the initial state $\ket{\Psi^{\mathcal{S}}_0}$ is a product state and invariant under arbitrary permutations of qudits in $\mathcal{H}_\mathcal{S}$, then the initial state $\ket{\rho_{\mathcal{S}}(0)\otimes  \rho_{\mathcal{S}}(0)}\rangle$, is invariant under permutation of qudits in $\widetilde{H}_{\mathcal{S}}$. As it will be clear from the subsequent discussion, this is also true for the evolved state $\ket{\rho_{\mathcal{S}}(t)\otimes  \rho_{\mathcal{S}}(t)}\rangle$: accordingly, the value of the purity $\mathcal{P}_{K}(t)$ only depends on the cardinality of $K$, $k=|K|$, and not on which sites belong to $K$ and we may write $\mathcal{P}_k(t) = \mathcal{P}_{K}(t)$.
	
	The formalism above allows  us  to write an equation describing  the evolution of the state $\mathbb{E}\left[\ket{\rho_{\mathcal{S}}(t)\otimes \rho_{\mathcal{S}}(t)}\rangle\right]$ under the continuous RUC introduced in Sec.~\ref{sec:model}. In particular, in the limit $M\to\infty$, we derive in Appendix~\ref{sec:lindbladian}
	\be
	\frac{\rm d}{{\rm d}t}\mathbb{E}\left[\ket{\rho_{\mathcal{S}}(t)\otimes \rho_{\mathcal{S}}(t)}\rangle\right]=-\mathcal{L}\mathbb{E}\left[\ket{\rho_{\mathcal{S}}(t)\otimes \rho_{\mathcal{S}}(t)}\rangle\right]\,,
	\label{eq:diff_eq}
	\ee
	where $\mathcal{L}$ is a super operator (the \emph{Lindbladian}) acting on $\widetilde{\mathcal{H}}_{\mathcal{S}}$, which reads
	\be
	\mathcal{L}=\frac{2\lambda_1}{N-1}\sum_{1\leq j<k\leq N}(1-\mathcal{U}_{j,k})+\frac{\lambda_2}{N}\sum_{j=1}^N \left(1- \ket{0,0,0,0}_j \bra{I^+}_j\right)\,,
	\label{eq:lindbladian_final}
	\ee
	with
	\be
	\mathcal{U}_{j,k}=\mathbb{E}\left[U^\ast_{j,k}\otimes U_{j,k}\otimes U^\ast_{j,k}\otimes U_{j,k}\right]\,.
	\label{eq:average_four_U}
	\ee
	
	In order to proceed further, we need to specify the probability distribution for the two-qudit unitary gates $U_{i,j}$, which in turn determines the average in Eq.~\eqref{eq:average_four_U}. As we already anticipated, we focus on two different physical situations. First we consider the case where $U_{i,j}$ are Haar-distributed over the group $U(d^2)$, which corresponds to a maximally chaotic evolution. Second, we consider random gates $U_{i,j}$ with a block structure determined by the presence of a $U(1)$ charge, as done for local RUCs in Refs.~\cite{rakovszky2018diffusive,khemani2018operator}. The two cases are treated separately in the next subsections.

	\subsection{Random Brownian circuit without conservation law}
	\label{sec:entropy_no_cons}

	As we have anticipated, we start  by choosing the unitary gates  $U_{i,j}$ to be Haar distributed over $U(d^2)$. In this case, the average in Eq.~\eqref{eq:average_four_U} can be computed easily, and we have (see for instance Refs.~\cite{nahum2018operator,vonKeyserlingk2018operator})
	\be
	\mathcal{U}_{j,k}=\frac{1}{d^4-1}\left[\ket{\mathcal{I}^{+}_{j,k}}\bra{\mathcal{I}^{+}_{j,k}}+\ket{\mathcal{I}^{-}_{j,k}}\bra{\mathcal{I}^{-}_{j,k}}-\frac{1}{d^2}\left(\ket{\mathcal{I}^{+}_{j,k}}\bra{\mathcal{I}^{-}_{j,k}}+\ket{\mathcal{I}^{-}_{j,k}}\bra{\mathcal{I}^{+}_{j,k}}
	\right)
	\right]\,,
	\label{eq:four_u_haar}
	\ee
	where
	\be
	\ket{\mathcal{I}^\pm_{j,k}}=\ket{I^\pm}_j\otimes \ket{I^\pm}_k\,.
	\ee
	Furthermore, throughout this section we initialize the system in the product state
	\be
	\ket{\Psi^{\mathcal{S}}_0}=\ket{1}^{\otimes N},
	\label{eq:initial_state}
	\ee
	With the above choices, one can now plug the explicit expression~\eqref{eq:four_u_haar} into \eqref{eq:lindbladian_final} and solve, at least numerically,  Eq.~\eqref{eq:diff_eq}.

	Unfortunately, the exact numerical solution to Eq.~\eqref{eq:diff_eq} is difficult to obtain for large values of $N$, as the dimension of $\widetilde{H}_{\mathcal{S}}$ grows exponentially with the system size. Luckily, in the present case the problem can be considerably simplified due to permutation symmetry between different qubits, and one does not need to solve Eq.~\eqref{eq:diff_eq} directly. Instead, based on Eq.~\eqref{eq:def_purity}, it is possible to derive the following system of differential equations
	\be
	\frac{{\rm d} \mathcal{P}_n(t)}{{\rm d} t}=\frac{2\lambda_1n(N-n) }{N-1}\frac{d}{(d^2+1)}\left[-\left(d+\frac{1}{d}\right)\mathcal{P}_n(t)+\mathcal{P}_{n-1}(t)+\mathcal{P}_{n+1}(t)\right]-\frac{\lambda_2n}{N}\left(\mathcal{P}_n-\mathcal{P}_{n-1}\right)\,,
	\label{eq:purtiy_equation}
	\ee
	for $n=0,\ldots ,N$ and with the convention $\mathcal{P}_{-1}(t)=\mathcal{P}_{N+1}(t)\equiv 0$. Here $\mathcal{P}_n(t)$ is the purity for a subsystem with $n$ qudits, while the initial conditions [corresponding to the state~\eqref{eq:initial_state}] are
	\be
	\mathcal{P}_{n}(0)=1\,,\qquad n=0,\ldots, N\,.
	\ee
	We note that Eq.~\eqref{eq:purtiy_equation} represents a rare example where an explicit result for the dynamics of the R\'enyi entropy can be obtained for open systems~\cite{alba2020spreading}. Since its derivation is rather technical, we reported it in Appendix~\ref{sec:purity_diff_eq}. 
	
	It is important to comment on this result. First, we note that setting $\lambda_2=0$ in Eq.~\eqref{eq:purtiy_equation}, we recover the same set of equations (up to prefactors) that was derived in Ref.~\cite{lashkari2013towards} for a Brownian Hamiltonian evolution. Thus, the internal dynamics driven by the RUC defined in Sec.~\ref{sec:model} is  qualitatively equivalent to a continuous Brownian Hamiltonian evolution. This observation allows us to apply directly some of the results of Ref.~\cite{lashkari2013towards} to our model.
	
	In particular, it was shown in  Ref.~\cite{lashkari2013towards} that the system~\eqref{eq:purtiy_equation} leads (for $\lambda_2=0$) to the emergence of a time scale which is logarithmic in $N$. More precisely, let us call $t^\ast_{(k)}$ the amount of time needed before the purity of a subsystem of size $k$ becomes less than $(1+\delta)2^{-k}$, where $\delta$ is a small positive real number, and $2^{-k}$ is the purity of a maximally mixed state. Then, for $0<\kappa<1$ fixed, it was shown that $t^\ast_{(\kappa N)} \sim \ln(N)t^\ast_{(1)}$. In our case, due to the choice made in Eq.~\eqref{eq:p_1}, we have that $t^\ast_{(1)}$ has a constant limit for $N\to\infty$, so that $t^\ast_{(\kappa N)} \sim \ln(N)$ for large $N$. In Ref.~\cite{lashkari2013towards} this was defined as the \emph{scrambling time} of the system. Note that, following later developments, the scrambling time is now usually defined as the time needed for OTOCs to decay to zero. However, the latter was shown to be also logarithmic in the system size $N$ for the Brownian Hamiltonian evolution of Ref.~\cite{lashkari2013towards}, see Ref.~\cite{zhou_operator_2019}, so that, up to prefactors, they can be identified in our model. 
	
\begin{figure}
		\begin{tabular}{ll}
			\hspace{-0.25cm}\includegraphics[width=0.485\textwidth]{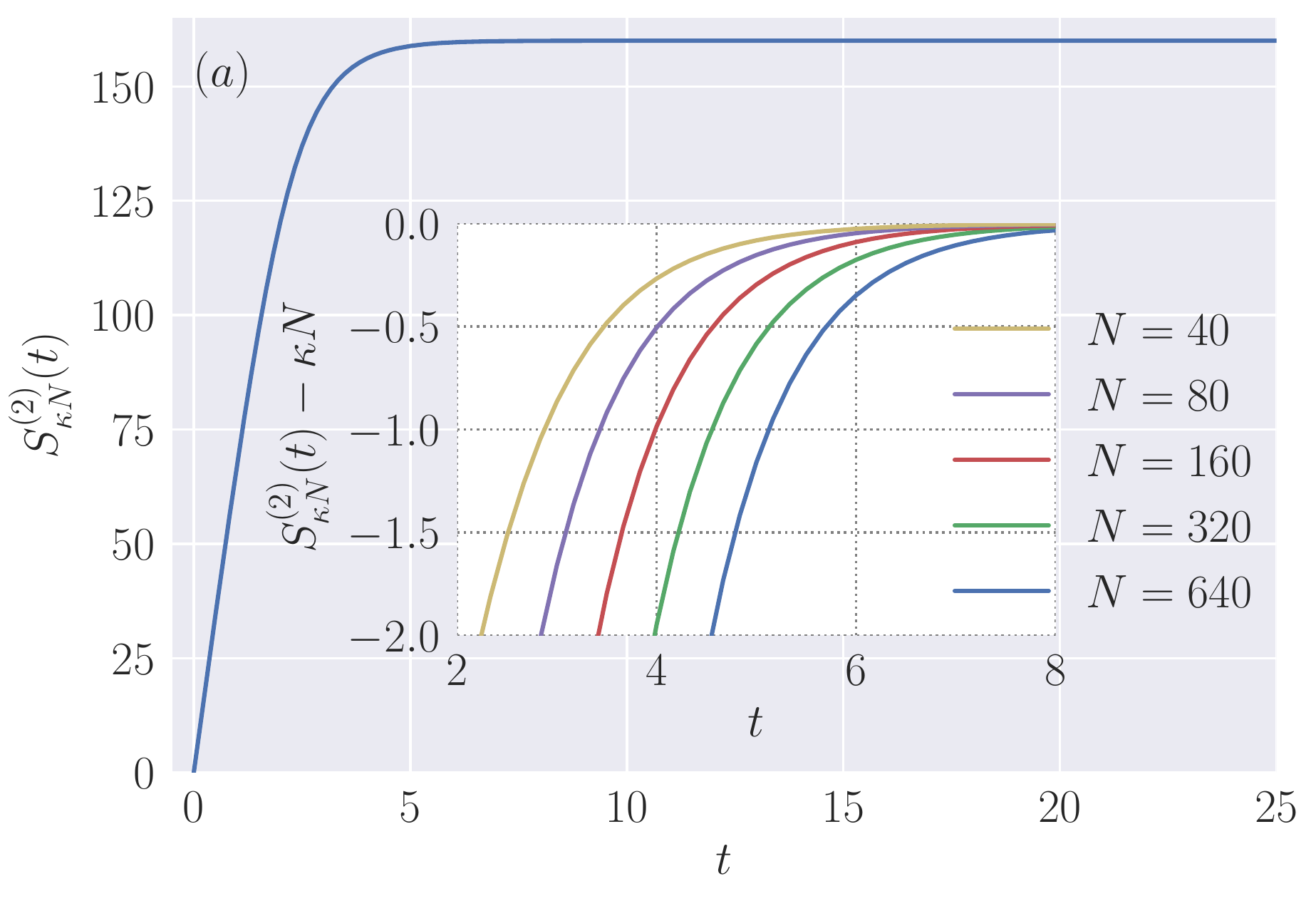} & \hspace{0.cm}\includegraphics[width=0.48\textwidth]{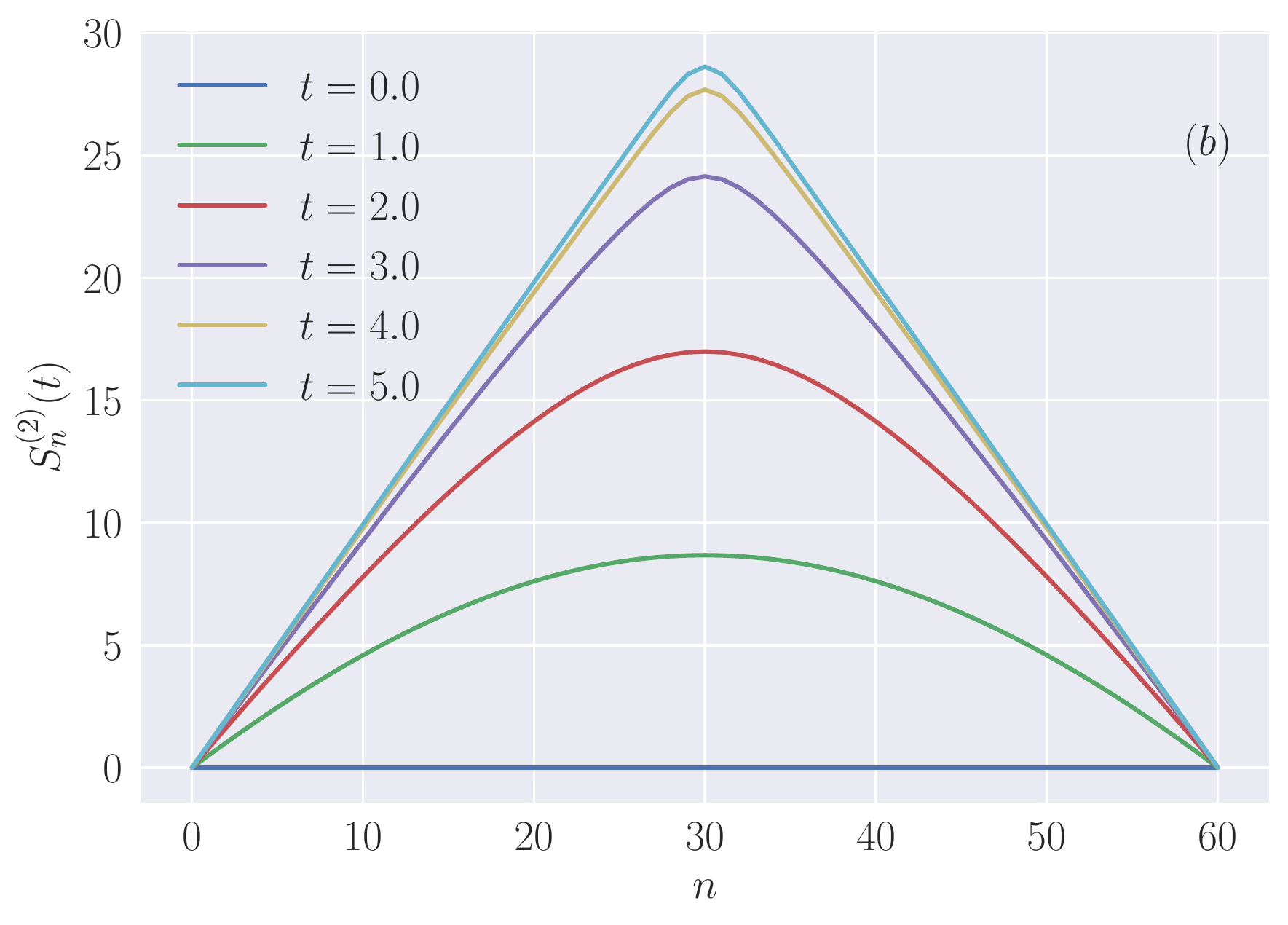}
		\end{tabular}
		\caption{R\'enyi entropy dynamics for different subsystems, obtained by solving Eq.~\eqref{eq:purtiy_equation} with $\lambda_2=0$ (no coupling with the environment), $\lambda_1=1$ and local dimension $d=2$. Subfigure $(a)$, main panel: R\'enyi-$2$ entropy $S^{(2)}_{\kappa N}(t)$ as a function of time, for $N=640$ and $\kappa=1/4$.  Inset: the plot shows the difference between $S^{(2)}_{\kappa N}(t)$ and its maximum possible value $\kappa N$ as a function of time and for increasing system sizes $N$. Denoting by $t^{\ast}_{(\kappa N)}$ the amount of time needed before $\mathcal{S}^{(2)}_{\kappa N}(t)-\kappa N$ becomes larger than a small negative constant $\varepsilon$ (cf. the main text), it is clear from the plot that $t^{\ast}_{(\kappa N)}\sim \ln (N)$. Subfigure $(b)$: R\'enyi-$2$ entropy $S^{(2)}_{n}(t)$ as a function of the subsystem size $n$, for different times.}
		\label{fig:free_case}
	\end{figure}

	The features of the entanglement dynamics for $\lambda_2=0$ discussed above are illustrated in Fig.~\ref{fig:free_case}, from which the emergence of a time scale logarithmic in $N$ is manifest.
	
	\begin{figure}
		\begin{tabular}{ll}
			\hspace{-0.25cm}\includegraphics[width=0.48\textwidth]{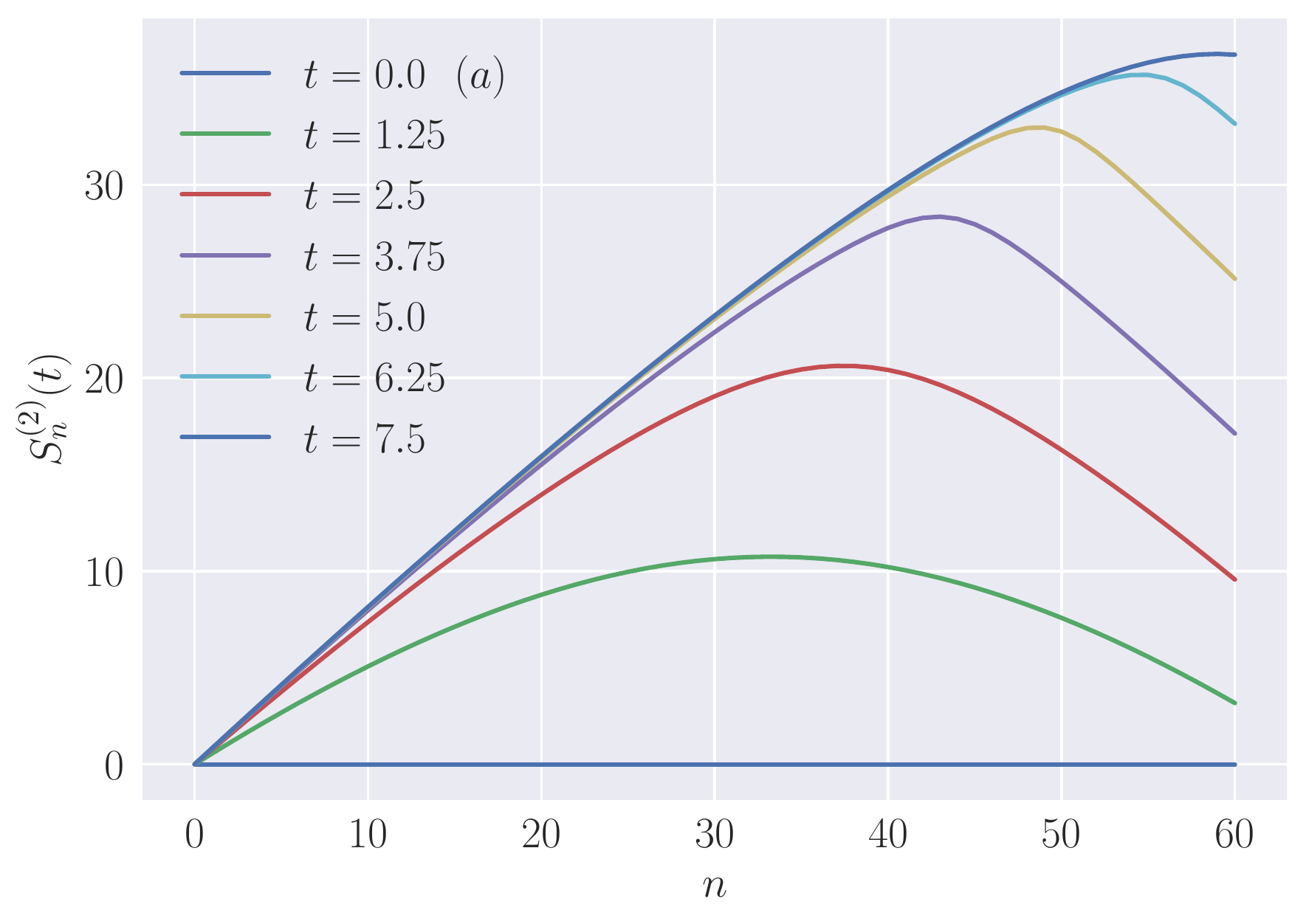} & \hspace{0.cm}\includegraphics[width=0.48\textwidth]{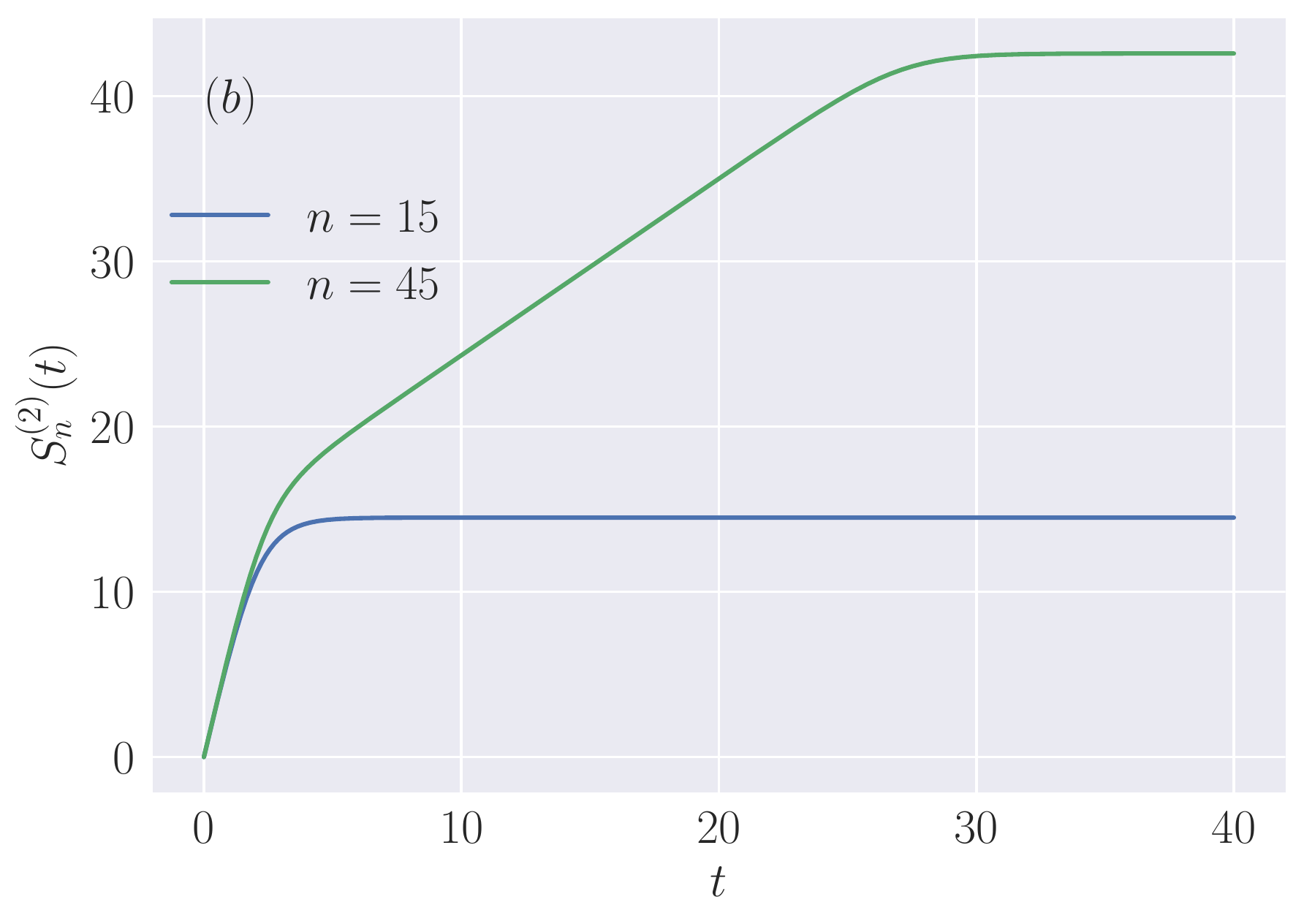}
		\end{tabular}
		\caption{R\'enyi entropy dynamics for different subsystems, obtained by solving Eq.~\eqref{eq:purtiy_equation}, for local dimension $d=2$ and $N=60$. Subfigure $(a)$: R\'enyi-$2$ entropy $S^{(2)}_{n}(t)$ as a function of the subsystem size $n$, for different times, with $\lambda_1=1$ and $\lambda_2=10$. Subfigure $(b)$: $S^{(2)}_{n}(t)$ as a function of time for subsystems containing $15$ and $N-15=45$ qudits. The parameters are chosen as $\lambda_1=1$, $\lambda_2=1.5$.}
		\label{fig:entropy_interact}
	\end{figure}
	
	Next, note that for $\lambda_1=0$ Eq.~\eqref{eq:purtiy_equation} predicts the purity of any subsystem to remain constant, namely $\mathcal{P}_n(t)\equiv 1$ for all values of $n$. This is due to the fact that, in each realization of the quantum circuit, $\mathcal{S}$ remains a pure state,  since the evolution only amounts to an exchange of qudits $\ket{1}$ and $\ket{0}$ between $\mathcal{S}$ and $\mathcal{E}$. 
	On the other hand, when both $\lambda_1,\lambda_2\neq 0$, the entanglement growth is non-trivial. In the following, we present our results based on the numerical solution of Eq.~\eqref{eq:purtiy_equation}.
	
	In Fig.~\ref{fig:entropy_interact}$(a)$ we report the numerical values of  $S^{(2)}_n(t)$ as a function of the subsystem size $n$, for different times $t$ and $\lambda_2\neq 0$. We can immediately appreciate that the effect of the environment is to increase the entanglement of $\mathcal{S}$, even though the environment itself consists of a product state. This is due to following mechanism: if $j$ is a qudit in $\mathcal{S}$, the internal dynamics will generate entanglement between $j$ and $\mathcal{S}\setminus j$. When $j$ is swapped with a qudit in $\mathcal{E}$, the latter becomes entanglement between $\mathcal{S}\setminus j$ and $\mathcal{\mathcal{E}}$. As a consequence, $\mathcal{S}$ does not remain in a pure state, and its entanglement grows in time. We also see that the R\'enyi entropy of $K$ and $\mathcal{S}\setminus K$ is not equal anymore, since the larger of the two can accommodate more entanglement with $\mathcal{E}$. 
	
	It is particularly interesting to follow the time evolution of a subsystem $K$ larger than half of the system size, as displayed in Fig.~\ref{fig:entropy_interact}$(b)$. We see that there are two relevant time scales that characterize its qualitative behavior: for short times, the R\'enyi entropy $S^{(2)}_{K}(t)$ is essentially on top of $S^{(2)}_{N\setminus K}(t)$. After a time $t_s$, $S^{(2)}_{K}(t)$ starts to increase with a constant slope up to a time $t_p$, at which saturation occurs (the indices $s$ and $p$ stand for ``scrambling'' and ``Page'' respectively: the use of these names will be justified in the next section). We can interpret the increase of $S^{(2)}_{K}(t)$ for $t<t_s$ as mainly due to the internal scrambling dynamics. Based on this picture, we expect $t_{s}\sim\ln(N)$, while, due to the normalization choice in Eqs.~\eqref{eq:p_1} and \eqref{eq:p_2}, $t_p\gg t_s$ for large $N$.
	
\begin{figure}
		\begin{tabular}{ll}
			\hspace{-0.25cm}\includegraphics[width=0.48\textwidth]{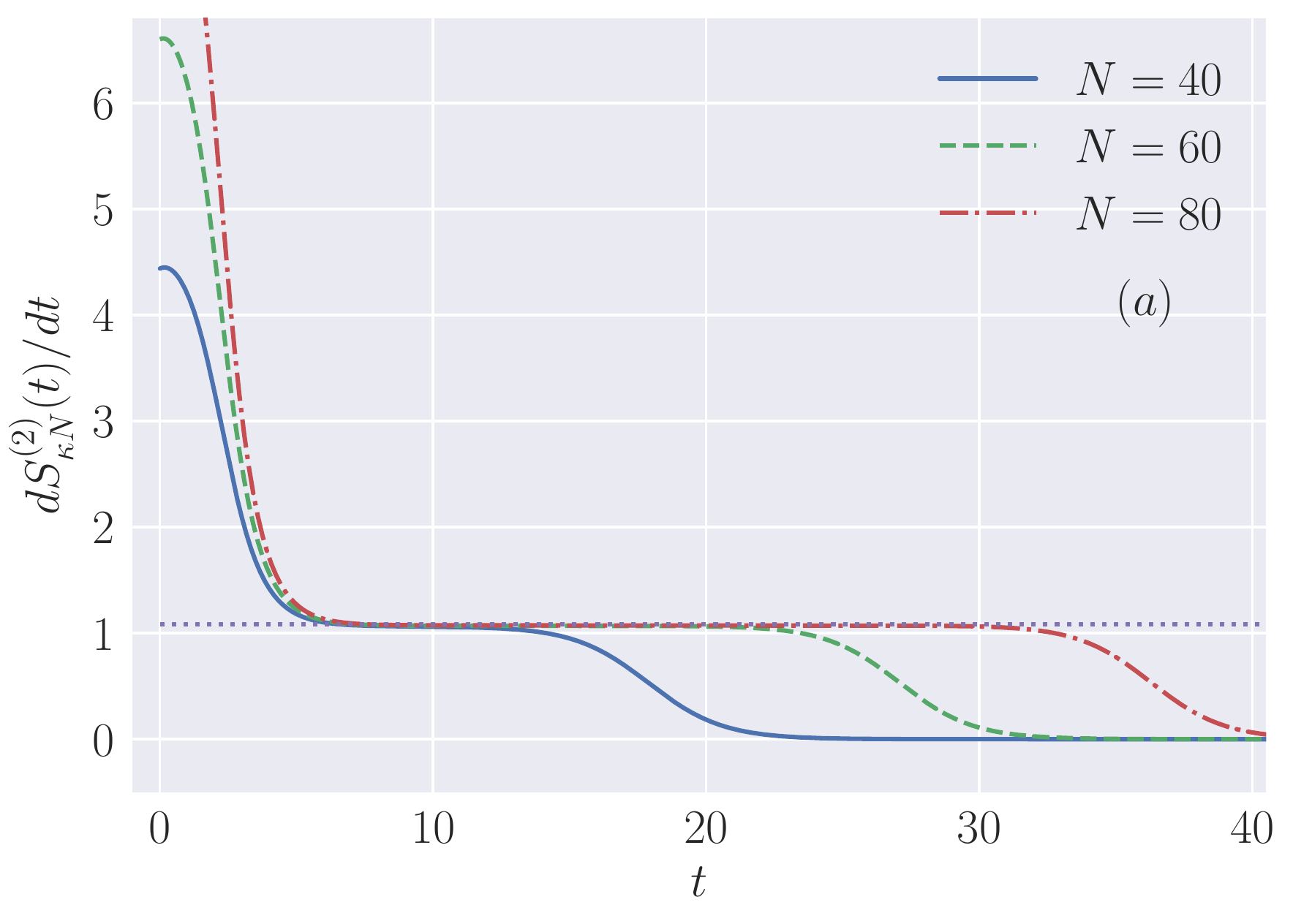} & \hspace{0.cm}\includegraphics[width=0.48\textwidth]{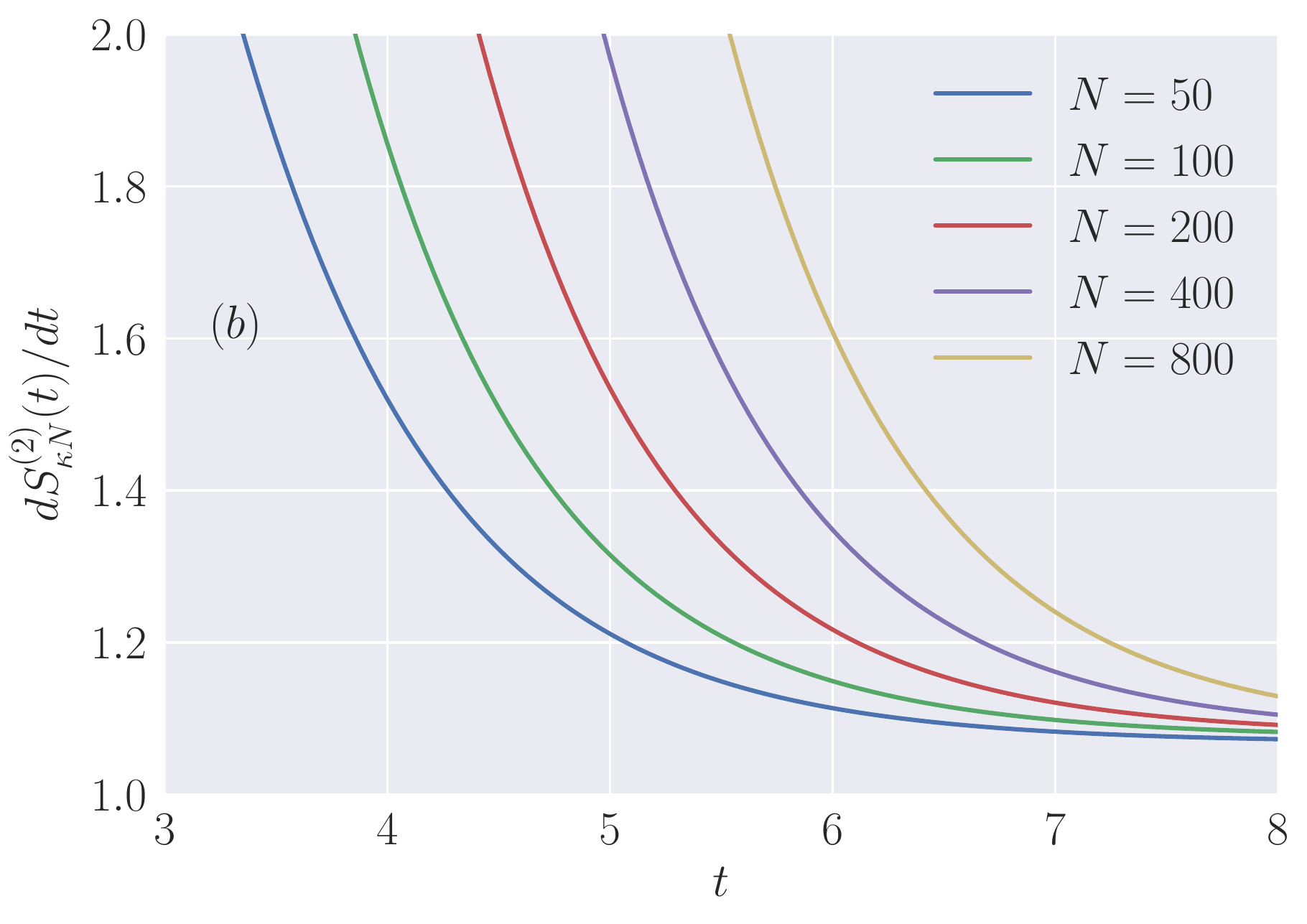}
		\end{tabular}
		\caption{Subfigure $(a)$:  time derivative of the R\'enyi-$2$ entropy $S^{(2)}_{\kappa N}(t)$ as a function of time, for different values of the system size $N$. The constant dotted line corresponds the value $s_{\lambda}$ defined in Eq.~\eqref{eq:s_lambda2}. Subfigure $(b)$: same data shown in the region close to $t_s$ (cf. the main text). In both figures we chose $\kappa=3/4$, local dimension $d=2$ and $\lambda_1=1$, $\lambda_2=1.5$. }
		\label{fig:entropy_derivative}
	\end{figure}

	To verify this, we have computed numerically the time derivative of $S^{(2)}_{K}(t)$, from which the emergence of different regimes is manifest, cf. Fig.~\ref{fig:entropy_derivative}. We see that for $t<t_s$ the derivative is large and increases with $N$, while for $t_s<t<t_p$ it approaches a constant $s_{\lambda_2}$ as $N\to\infty$. It is not straightforward to compute $s_{\lambda_2}$ directly from Eq.~\eqref{eq:purtiy_equation}: indeed, while at short times the r.h.s of Eq.~\eqref{eq:purtiy_equation} is dominated by the term proportional to $\lambda_1$, for $t\sim t_s$ the absolute value of the latter becomes comparable to the term proportional to $\lambda_2$ and both contribute in a non-negligible way to $s_{\lambda_2}$. Nevertheless, we can make the conjecture
	\be
	s_{\lambda_2}=\frac{(d-1)}{d}\frac{\lambda_2}{\ln (2)}\,.
	\label{eq:s_lambda2}
	\ee
	In order to motivate Eq.~\eqref{eq:s_lambda2}, we consider the case $K=\mathcal{S}$, so that only the term proportional to $\lambda_2$ in Eq.~\eqref{eq:purtiy_equation} is non-vanishing. In the limit $\lambda_2\to 0$,  one can make the assumption  that that after a time $t>t_{s}\sim \ln(N)$ the  system is almost maximally scrambled. Then, for large $n$ one would get $\mathcal{P}_n(t)\simeq d^{-N+n}$, so that
	\be
	\frac{{\rm d} \mathcal{P}_N(t)}{{\rm d} t}\simeq -\frac{\lambda_2(d-1)}{d}\mathcal{P}_N(t)\,,
	\ee 
	and so
	\be
	\frac{\rm d}{{\rm d}t}S^{(2)}_N(t)=-\frac{\rm d}{{\rm d}t}\log_{2}\mathcal{P}_N(t)=-\frac{1}{\mathcal{P}_N(t)\ln 2 }\frac{\rm d}{{\rm d}t}\mathcal{P}_N(t)=\frac{(d-1)}{d}\frac{\lambda_2}{\ln 2}\,.
	\ee
	Remarkably, we found that Eq.~\eqref{eq:s_lambda2} is in perfect agreement with the numerical solution to Eq.~\eqref{eq:purtiy_equation} for arbitrary values of $\lambda_1$  and $\lambda_2$, and also for general $K\subset \mathcal{S}$ (with $|K|>N/2$),  suggesting  that it should be possible to derive it rigorously from Eq.~\eqref{eq:purtiy_equation}. 
	
	We can estimate $t_s$ precisely, by defining it as the amount of time needed in order for $d S^{(2)}_{K}(t)/dt$ to become smaller than $s_{\lambda_2}+\varepsilon$, where $\varepsilon$ is a positive small number. We see clearly from Fig.~\ref{fig:entropy_derivative}$(b)$ that $t_s\sim \ln(N)$, as we also verified with a quantitative fit. On the other hand, one can define analogously $t_p$ to be the amount of time needed in order for $d S^{(2)}_{K}(t)/dt$ to be smaller than a small positive constant, and as it is clear from Fig.~\ref{fig:entropy_derivative}$(a)$, one has $t_{p}\sim N$, so that indeed $t_p\gg t_s$.

	In summary, the above analysis shows that in the presence of both internal dynamics and system-environment interaction, two distinct time scales emerge: one can be associated with the internal scrambling time $t_s$, with $t_s\sim \ln (N)$, while the other, $t_{p}$, depends on the interaction with $\mathcal{E}$, and for the RUC constructed in Sec.~\ref{sec:model} we have  $t_{p}\sim N$.

	\subsection{Random dynamics with a conserved $U(1)$ charge}
	\label{sec:entropy_u1}
	
	In the previous subsection we have seen that the second R\'enyi entropy for a  subsystem $K\subset\mathcal{S}$ grows always monotonically with time, even if $\mathcal{E}$ is initialized in the product state $\ket{0}^{\otimes M}$. On the other hand, in a unitary black hole evaporation process, one expects that the entanglement follows a ``Page-like'' behavior in time~\cite{page2005hawking}: namely it initially grows but starts to decrease in the middle stage of the evaporation, and eventually vanishes when the black hole evaporates completely~\footnote{In some toy models such as Ref. \cite{agarwal_toy_2019}, this was introduced by hand, moving qubits from black hole to the bath at each time step.}. This difference between a black hole and random tensor networks originates from the absence of energy conservation in the latter. In the long-time limit, the black hole returns to a vacuum state since its energy leaves with the radiation, while the random tensor network model approaches a random state with large  entanglement entropy between the system and the bath.
	
	It is difficult to introduce energy conservation in tensor network models, but it is possible to introduce a $U(1)$ charge conservation, which plays a similar role. When the bath is infinitely large and initialized in the zero ({\it i.e.} lowest ) charge ``vacuum" state, the black hole charge will gradually decrease and approach zero in the final state. As long as the zero-charge state is unique, the black hole entropy will eventually vanish in the long-time limit (for a very different approach that achieves similar phenomena, see Ref.~\cite{czech_black_2011}).
	
	We implement a dynamics with a $U(1)$ conserved charge by imposing that the two-qudit unitary gates $U_{i,j}$ have some special structure, as done in Refs.~\cite{rakovszky2018diffusive,khemani2018operator} for the case of spatially local circuits. For the rest of this section, we will focus on the case of qubits, namely $d=2$. Then, following~\cite{rakovszky2018diffusive,khemani2018operator} we consider gates of the form
	\be
	U_{i,j}=
	\begin{pmatrix}
		U_{Q=0} & & \\
		& U_{Q=1} & \\
		& & U_{Q=2}\\
	\end{pmatrix}\,,
	\label{eq:conservation_u}
	\ee
	where the first and last blocks are $1 \times 1$ and the second block is a $2 \times 2$ Haar-random unitary matrix. Since the interaction with the environment is driven by swap gates, Eq.~\eqref{eq:conservation_u} defines a dynamics conserving the charge
	\be
	Q_{\mathcal{S}\cup \mathcal{E}}=\sum_{j\in\mathcal{S}\cup\mathcal{E}}q_{j}\,,
	\ee
	where the charge operator is
	\be
	q_j=
	\begin{pmatrix}
		0 & 0\\
		0 & 1
	\end{pmatrix}\,.
	\ee
	on each site. 
	
	Averaging over unitary gates of the form~\eqref{eq:conservation_u} introduces additional computational difficulties with respect to the case of Haar-distributed operators. In particular, by exploiting the results derived in Ref.~\cite{rakovszky2018diffusive}, Eq.~\eqref{eq:four_u_haar} has to be replaced by
	\be
	\begin{aligned}
		\mathcal{U}_{j,k}=& \sum_{s=\pm} \sum_{Q_{j} \neq Q_{k}} \frac{1}{d_{Q_{j}} d_{Q_{k}}}\left|\mathcal{I}_{Q_{j} Q_{k}}^{s}\right\rangle\left\langle\mathcal{I}_{Q_{j} Q_{k}}^{s}\right| \\ &+\sum_{s=\pm} \sum_{\mathcal{Q}} \frac{1}{d_{Q}^{2}-1}\left[\left|\mathcal{I}_{Q Q}^{s}\right\rangle\bra{\mathcal{I}_{Q Q}^{s}}-\frac{1}{d_{Q}} \ket{\mathcal{I}_{Q Q}^{s}}\bra{\mathcal{I}_{Q Q}^{-s}}\right]\,.
	\end{aligned}
	\label{eq:four_u_conserved}
	\ee
	Here $\ket{\mathcal{I}_{Q_{j} Q_{k}}^{\pm}}$ are states living in the tensor-product of two local sites of the four replica space. In terms of single-site states, they can be written as
	\bea
	\left|\mathcal{I}_{Q_{j} Q_{k}}^{+}\right\rangle&=&\sum_{\alpha \beta \gamma \delta}|\alpha \alpha \beta \beta\rangle_{j}|\gamma \gamma \delta \delta\rangle_{k} \delta_{\alpha+\gamma=Q_{j}} \delta_{\beta+\delta=Q_{k}}\,,\\
	\left|\mathcal{I}^-_{Q_{j} Q_{k}}\right\rangle&=&\sum_{\alpha \beta \gamma \delta}|\alpha \beta \beta \alpha\rangle_{j}|\gamma \delta \delta \gamma\rangle_{k} \delta_{\alpha+\gamma=Q_{1}} \delta_{\beta+\delta=Q_{k}}\,,
	\eea
	where, for the case of qubits, the Greek indices take value in $\{0,1\}$, while $d_{0}=d_2=1$ and $d_1=2$.
	
	The form of Eq.~\eqref{eq:four_u_conserved} makes the computations considerably more involved. In particular, one can not derive a set of $N+1$ differential equations for the purity, and a different strategy is needed to obtain $S^{(2)}_K(t)$ efficiently. Luckily, one can exploit an observation of Ref.~\cite{rakovszky2018diffusive}. Namely, the states $\ket{\mathcal{I}_{Q_{j} Q_{k}}^{\pm}}$ can be written in terms of the following $6$ states~\cite{rakovszky2018diffusive}
	\bea
	\ket{\mathbf{0}} &\equiv&\ket{0000}\,, \qquad \ket{\mathbf{1}}  \equiv \ket{1111}\,, \label{eq:basis_1}\\ 
	\ket{A} &\equiv& \ket{1100}\,,\qquad \ket{B}  \equiv \ket{0011}\,, \\ \label{eq:basis_2}
	\ket{C} &\equiv&\ket{1001}\,,\qquad \ket{D}  \equiv \ket{0110} \,. \label{eq:basis_3}
	\eea
	Once again, we stress that the states $\ket{\mathbf{0}}$, $\ket{\mathbf{1}}$, $\ket{A}$, $\ket{B}$, $\ket{C}$ and $\ket{D}$ live in a single local space of the four replicas. This means that the evolution dictated by the averaged gates $\mathcal{U}_{j,k}$ effectively takes place in a Hilbert space $\widetilde{\mathcal{H}}^{\rm eff}_{\mathcal{S}}=h^{\rm eff}_1\otimes \cdots \otimes {h}^{\rm eff}_N$, where ${h}^{\rm eff}_j\simeq \mathbb{C}^6$ so that $\mathcal{U}_{i,j}$ is a matrix acting on the space $\mathbb{C}^6\otimes \mathbb{C}^6$. 
	
	The above consideration becomes particularly powerful when combined with the underlying permutational symmetry of the operator $\mathcal{L}$ and of the initial state $\ket{\rho_{\mathcal{S}}(0)\otimes \rho_{\mathcal{S}}(0)}\rangle$. Indeed, this allows us to exploit a logic which is similar to the one developed in Ref.~\cite{sunderhauf_quantum_2019}, and obtain an efficient scheme to compute the evolution of the system in a numerically exact fashion.
	
	We start by introducing the following class of permutationally invariant states on the space $\widetilde{\mathcal{H}}^{\rm eff}_{\mathcal{S}}$
	\bea
	\ket{n_{\mathbf0},n_{\mathbf1},n_A,n_B,n_C,n_D}&=&\frac{1}{\sqrt{N!n_{\mathbf0}!n_{\mathbf1}!n_A!n_B!n_C!n_D!}}\sum_{\pi\in S_N}\pi \underbrace{\ket{\mathbf 0}\otimes \cdots \otimes \ket{\mathbf0}}_{n_{\mathbf0}}\otimes\cdots \otimes  \underbrace{\ket{D}\otimes \cdots \otimes \ket{D}}_{n_D} \pi^{-1}\,.
	\eea
	Importantly, we can rewrite these states by introducing a set of bosonic creation operators as~\cite{sunderhauf_quantum_2019}
	\be
	\ket{n_{\mathbf0},n_{\mathbf{1}},n_A,n_B,n_C,n_D}=\frac{1}{\sqrt{n_{\mathbf0}!n_{\mathbf1}!n_A!n_B!n_C!n_D!}}\left(a_{\mathbf0}^{\dagger}\right)^{n_{\mathbf0}}\left(a_{\mathbf1}^{\dagger}\right)^{n_{\mathbf1}}\left(a_{A}^{\dagger}\right)^{n_A}\left(a_{B}^{\dagger}\right)^{n_B}\left(a_{C}^{\dagger}\right)^{n_C}\left(a_{D}^{\dagger}\right)^{n_D}\ket{\Omega}\,,
	\label{eq:bosonic_basis_states}
	\ee
	where $\left[a_{j},a_{k}\right]=\left[a^{\dagger}_{j},a^{\dagger}_{k}\right]=0$ and $\left[a_{j},a^{\dagger}_{k}\right]=\delta_{j,k}$, while $\ket{\Omega}$ is a vacuum state. One of the advantages of the bosonic representation is that the operator $\mathcal{L}$, the initial state $\ket{\rho_{\mathcal{S}}(0)\otimes \rho_{\mathcal{S}}(0) }\rangle$ and the vector $\ket{W_K}\rangle$ defined in Eq.~\eqref{eq:vector_form} admit a simple expression in terms of the $a$-operators. Since we won't make use of them in the following, we report them in Appendix~\ref{sec:bosonic_modes}, to which we refer the interested reader.
	
	\begin{figure}
		\begin{tabular}{ll}
			\hspace{-0.25cm}\includegraphics[width=0.48\textwidth]{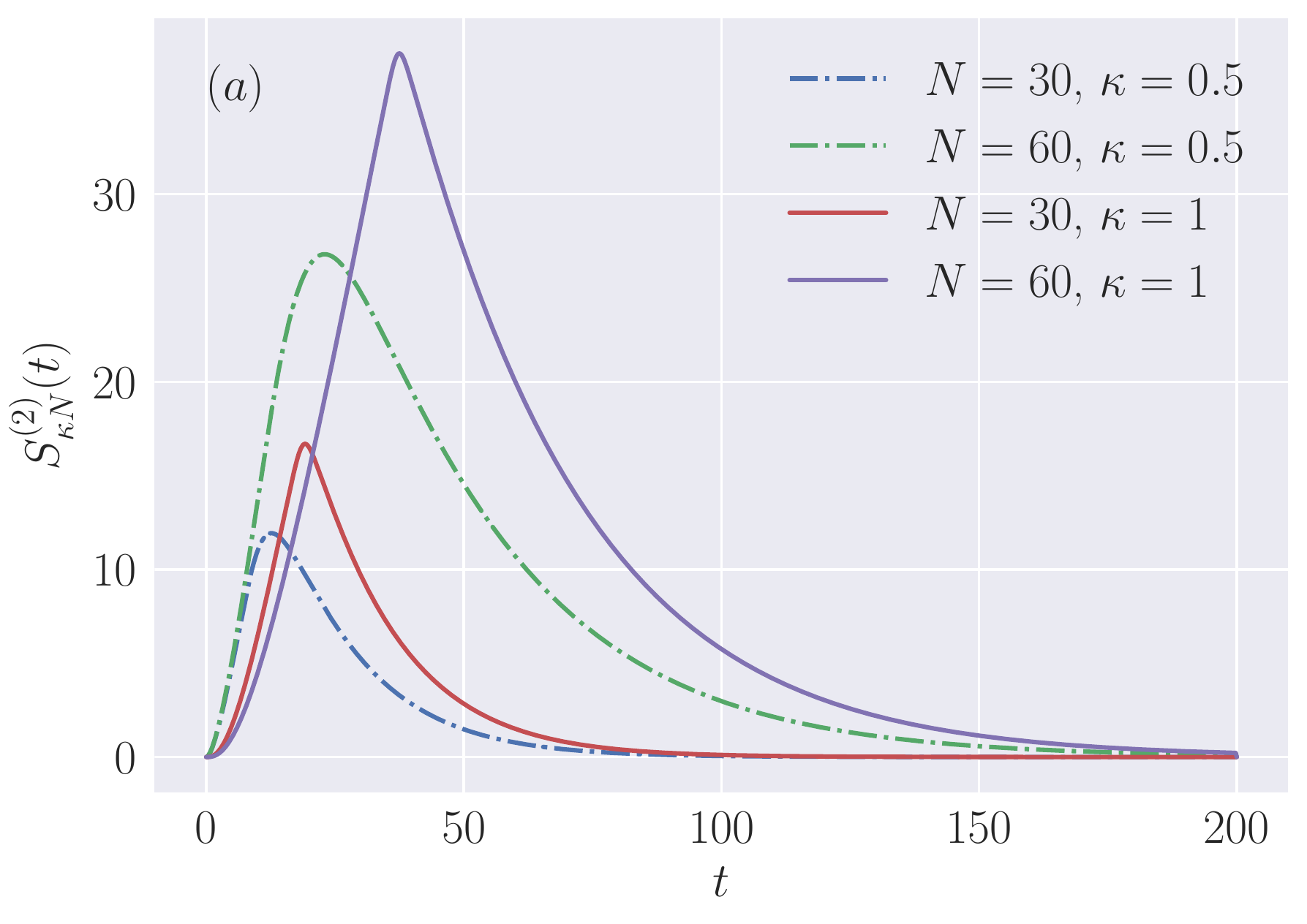} & \hspace{0.cm}\includegraphics[width=0.48\textwidth]{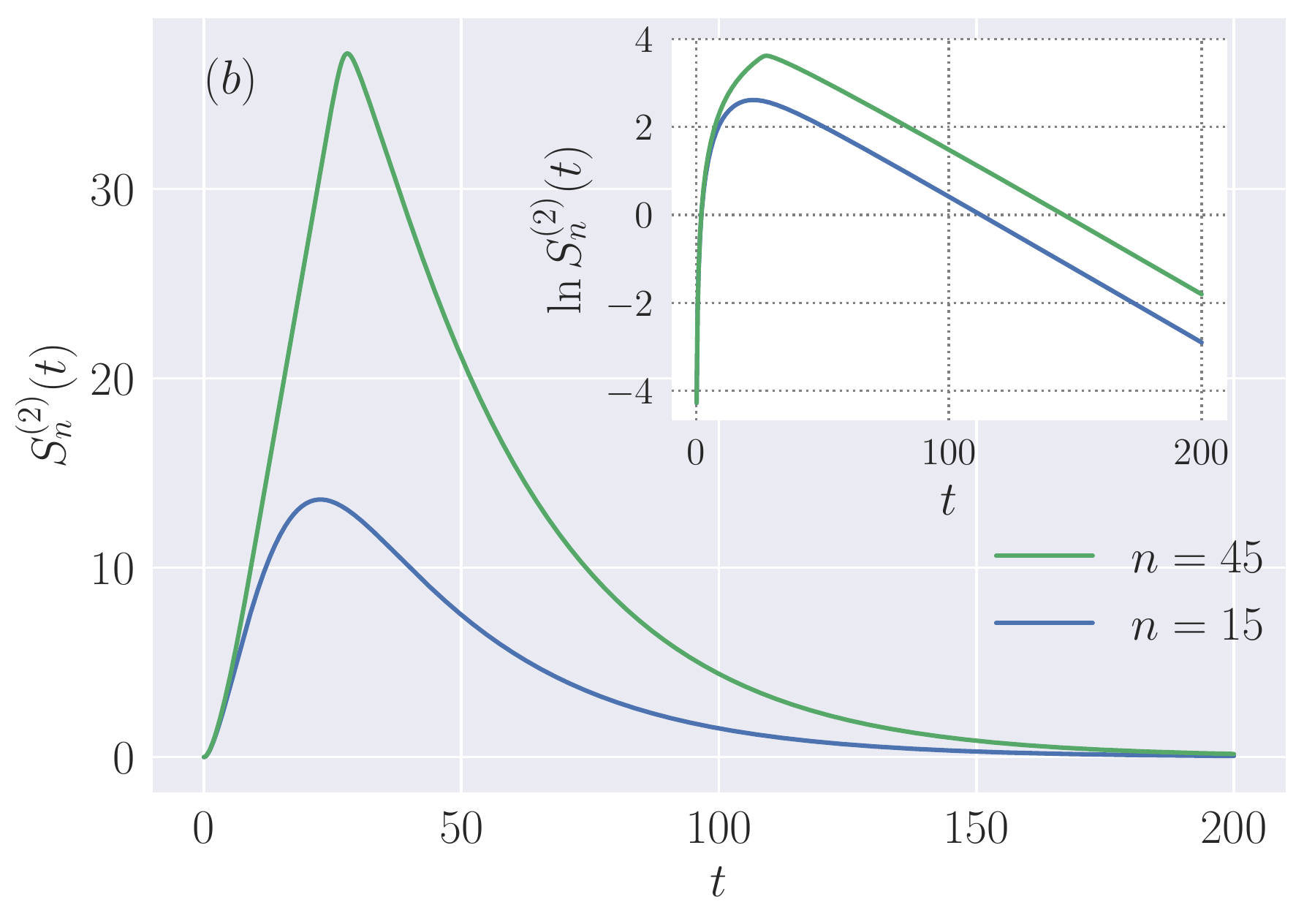}
		\end{tabular}
		\caption{R\'enyi entropy $S^{(2)}_{n}(t)$ as a function of time for $\lambda_1=1$ and $\lambda_2=2$. The plots correspond to a random evolution in the presence of a global $U(1)$ conserved charge, while the system is initialized as in Eq.~\eqref{eq:initial_state}. Subfigures $(a)$: $S^{(2)}_{\kappa N}(t)$  for different values of the system size $N$ and $\kappa=0.5, 1$. Subfigure $(b)$: $S^{(2)}_{n}(t)$ for $N=60$ and subsystems containing $15$ and $N-15=45$ qubits. Inset:  $\ln S^{(2)}_{n}(t)$ for the same values of $n$.}
		\label{fig:entropy_conserved_prod}
	\end{figure}
	
	Since both the initial state and the Lindbladian $\mathcal{L}$ are invariant under arbitrary permutation of qubits, the states~\eqref{eq:bosonic_basis_states} form a basis of the Hilbert space in which the dynamics takes place. Crucially, the corresponding dimension is
	\be
	D_{\rm perm}=\binom{N+5}{5}=N^{5}+O(N^{4})\,,
	\ee 
	and thus grows only polynomially (rather than exponentially) with $N$. In practice $D_{\rm perm}$ is still very large for the values of $N$ considered in the previous subsection. Nevertheless, we were able to perform numerically exact calculations up to $N=80$. This was done by implementing the matrix  corresponding to $\mathcal{L}$ in the vector basis~\eqref{eq:bosonic_basis_states}, and then computing the evolved state $\ket{\rho_{\mathcal{S}}(t)\otimes \rho_{\mathcal{S}}(t)}\rangle$ solving the system of differential equations encoded in Eq.~\eqref{eq:diff_eq}. Note that in this way we did not need to diagonalize exactly the matrix associated with $\mathcal{L}$, which would be unfeasible for $N=80$. In  rest of this section, we report the numerical results obtained by following  the above procedure.

	\begin{figure}
		\begin{tabular}{ll}
			\hspace{-0.25cm}\includegraphics[width=0.48\textwidth]{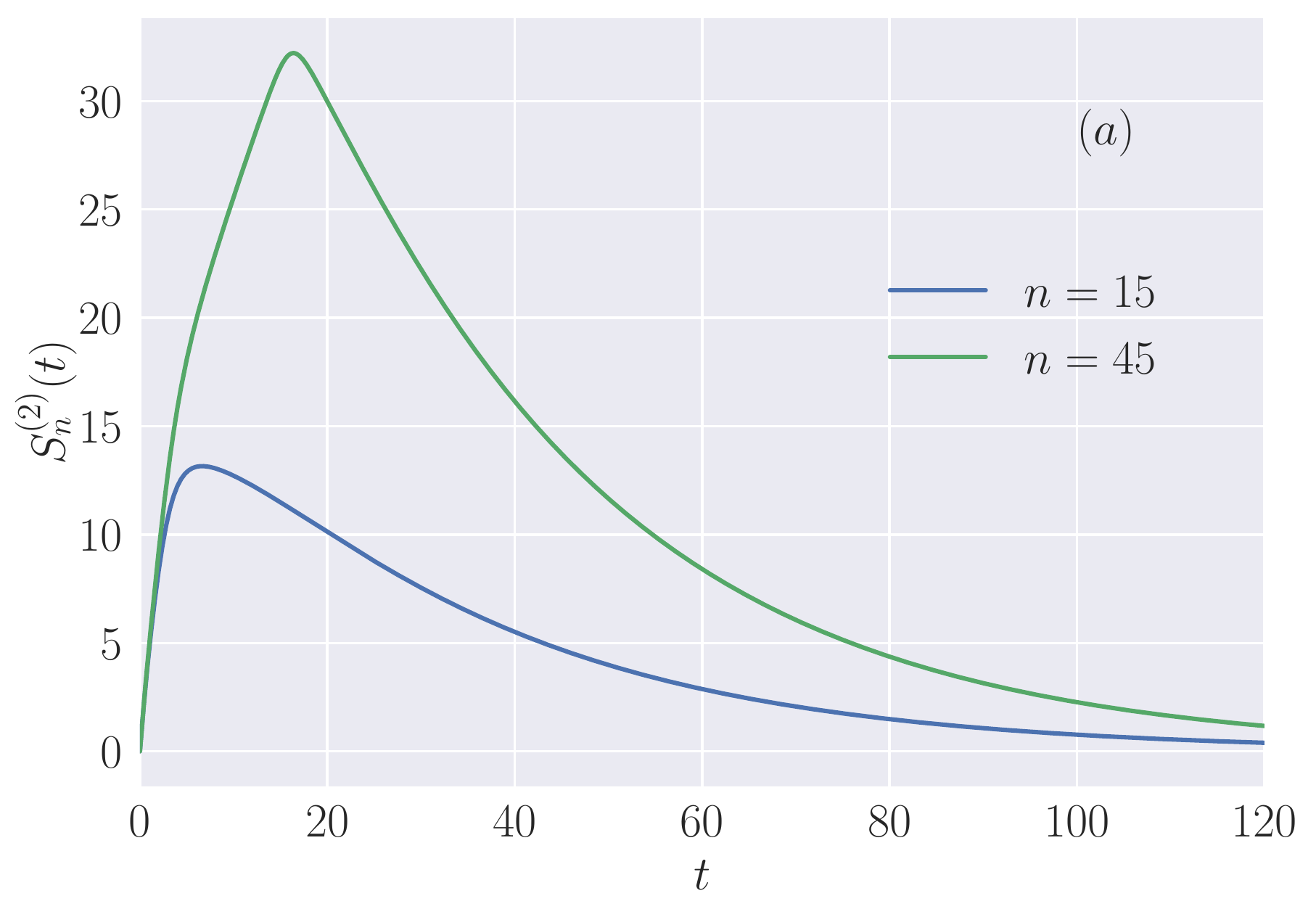} & \hspace{0.cm}\includegraphics[width=0.48\textwidth]{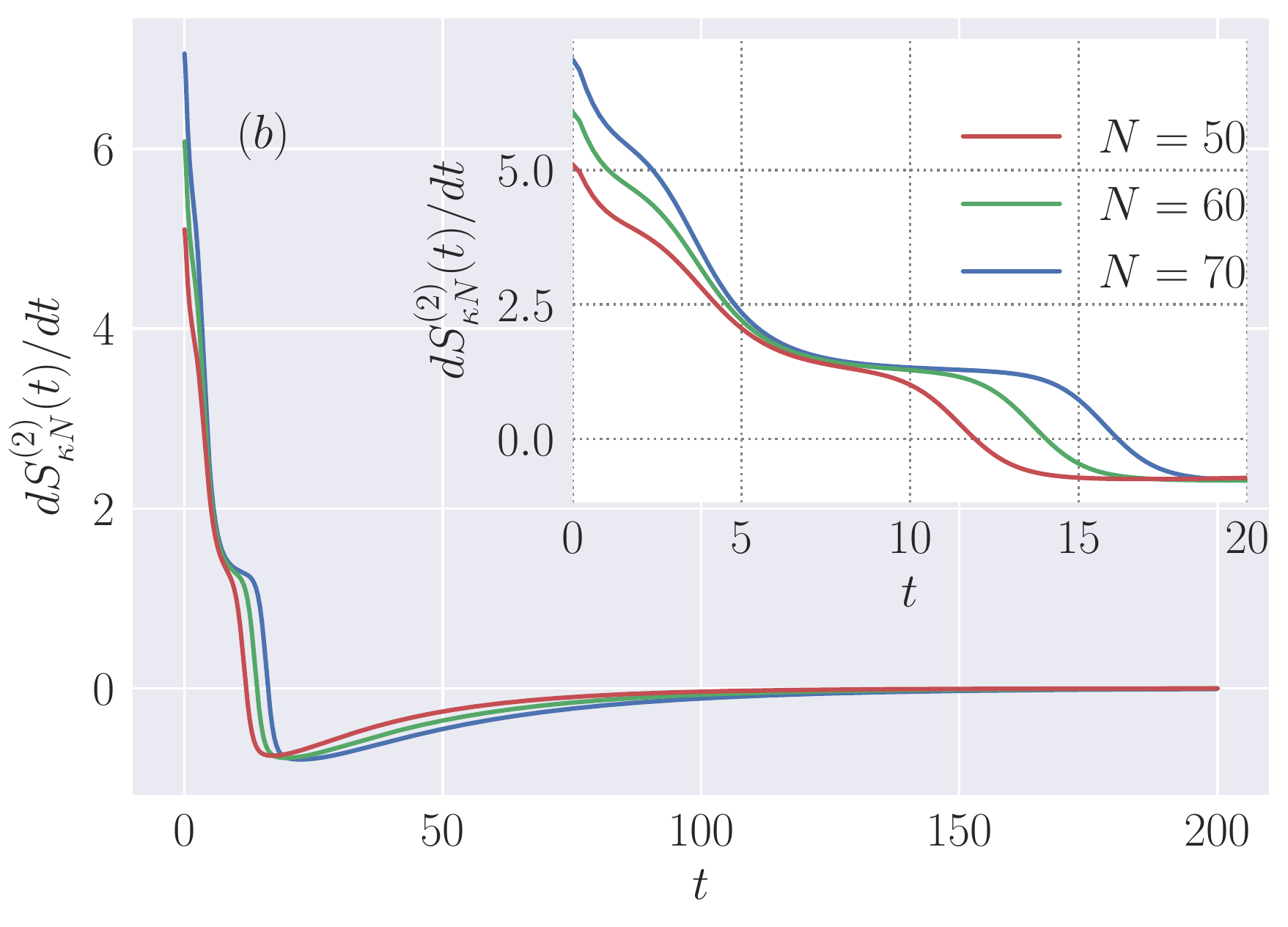}
		\end{tabular}
		\caption{Subfigure $(a)$: R\'enyi entropy $S^{(2)}_{n}(t)$ as a function of time for $N=60$ and subsystems containing $15$ and $N-15=45$ qubits. The plot corresponds to a random evolution in the presence of a global $U(1)$ conserved charge, while the system is initialized as in Eq.~\eqref{eq:initial_state_halfhalf}. The evolution parameters are set to $\lambda_1=1$ and $\lambda_2=2$. Subfigure $(b)$: time derivative of the R\'enyi-$2$ entropy $S^{(2)}_{\kappa N}(t)$ as a function of time, for $\kappa= 7/10$ and increasing values of $N$. The evolution parameters are set to $\lambda_1=1$ and $\lambda_2=2 $. The inset shows the same plot in the time region close to $t_s$ (cf. the main text).}
		\label{fig:entropy_conserved_half_half}
	\end{figure}
	
	We first consider the case where the system is initialized in the product state~\eqref{eq:initial_state}, and study the time evolution of the R\'enyi entropy $S^{(2)}_{K}(t)$ for different subsystems $K\subset \mathcal{S}$, as reported in Fig.~\ref{fig:entropy_conserved_prod}. We immediately see that the qualitative behavior is different from the Haar-scrambled case, since $S^{(2)}_{K}(t)$ is a non-monotonic function. For a given subsystem $K$, with $|K|=\kappa N$ and $\kappa\in [0,1]$, we have  verified that the time $t_p$ at which $S^{(2)}_{\kappa N}(t)$ reaches its maximum grows linearly with  $N/\lambda_2$. We call  $t_p$ the Page-time~\cite{page2005hawking}. After $t_p$, we see from Fig.~\ref{fig:entropy_conserved_prod} that $S^{(2)}_{\kappa N}(t)$ decreases and approaches zero exponentially fast, with an exponent that does not appear to depend on $\kappa$.

	Besides its non-monotonic behavior,  $S^{(2)}_{K}(t)$ displays another qualitative difference. Indeed, the initial state of $\mathcal{S}$, defined  in  Eq.~\eqref{eq:initial_state}, is a fixed point for the internal dynamics. Hence, at short times, one can not distinguish clearly the contribution of  the internal scrambling, since the initial growth of $S^{(2)}_{K}(t)$  is only due to system-environment coupling (there is no evolution if $\lambda_2=0$). For  this reason,  we consider in Fig.~\ref{fig:entropy_conserved_half_half} the R\'enyi entropy $S^{(2)}_{K}(t)$, for the initial state
	\be
	\ket{\Psi^{\mathcal{S}}_0}=\bigotimes_{j=1}^{N/2}\ket{0}_j \bigotimes_{j=N/2+1}^N \ket{1}_j\,. 
	\label{eq:initial_state_halfhalf}
	\ee
	Note that this state is not invariant under permutation of qubits.  Accordingly, we consider a protocol where not only  we sample over different realizations of  the RUC, but we also take an average over all the initial product states obtained by permuting the qubits in ~\eqref{eq:initial_state_halfhalf}: namely, we average over all product states where half of the qubits  are initialized to $\ket{0}$, and the rest are set to $\ket{1}$. It is straightforward to see that in the four replica space $\widetilde{H}_{\mathcal{S}}$ the state $\ket{\rho_{\mathcal{S}}(0)\otimes\rho_{\mathcal{S}}(0)}\rangle$ (obtained after averaging over the initial configurations) is indeed permutationally invariant, and we can employ the approach explained above.
	
	As expected, we see from  Fig.~\ref{fig:entropy_conserved_half_half}$(a)$ a separation of time scales for the initial state~\eqref{eq:initial_state_halfhalf}. In order to make  this more transparent, and  following the previous  section, we report in  Fig.~\ref{fig:entropy_conserved_half_half}$(b)$  the time derivative of the R\'enyi-$2$ entropy $S^{(2)}_{\kappa N}(t)$. Although the results are now plagued by larger finite-$N$ effects, we can see the same qualitative behavior displayed in Fig.~\ref{fig:entropy_derivative} for the Haar-scrambled dynamics. In  particular, after a time $t_s \sim \ln N$ the derivatives appear to approach a plateau, and it remains approximately constant for $t_{ s}<t<t_{p}$, where $t_p\sim N$ is the Page time.  
			
\begin{figure}
		\begin{tabular}{ll}
			\hspace{-0.25cm}\includegraphics[width=0.48\textwidth]{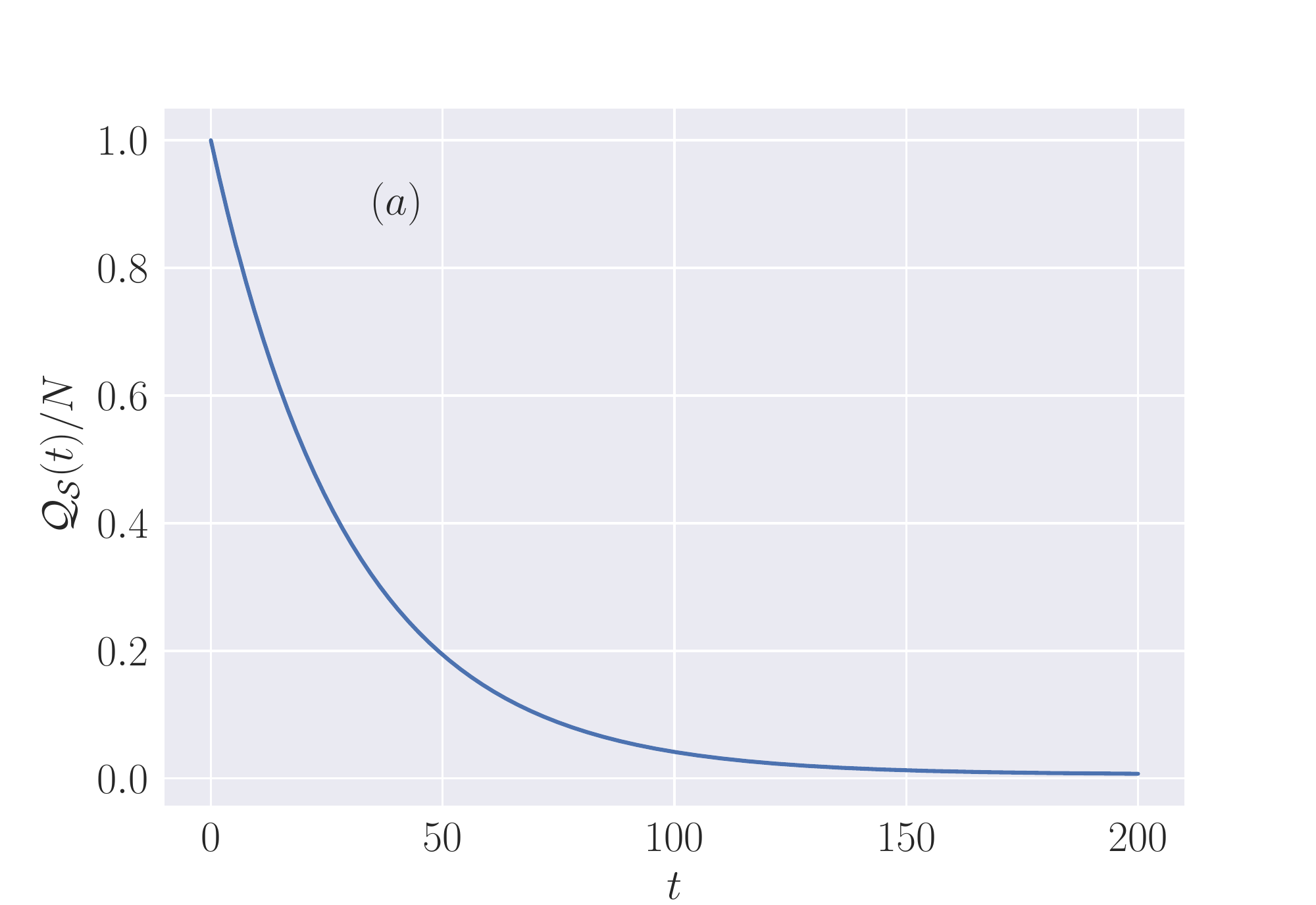} & \hspace{0.cm}\includegraphics[width=0.48\textwidth]{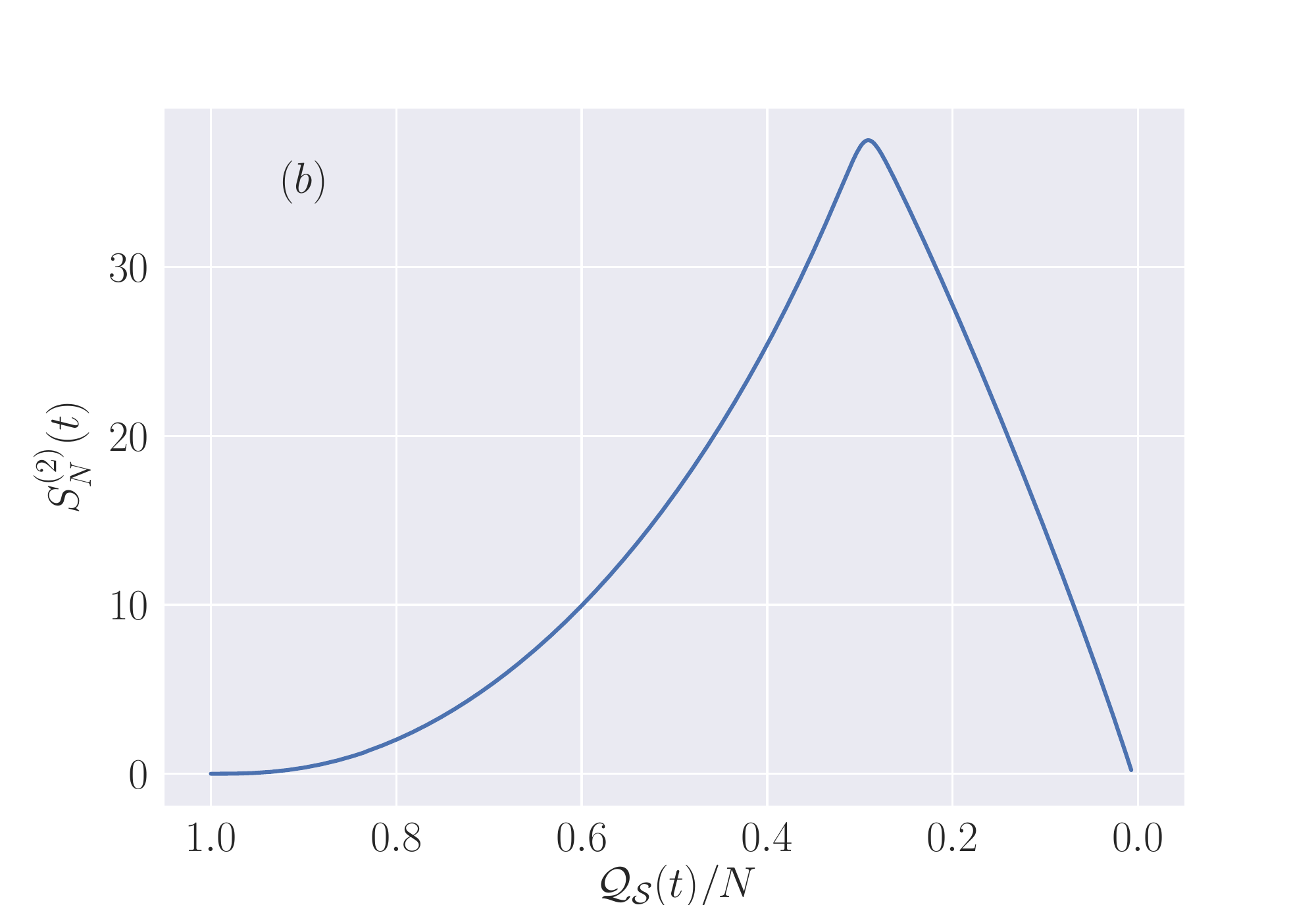}
		\end{tabular}
		\caption{Subfigure $(a)$: time evolution of the system charge $\mathcal{Q}_{\mathcal{S}}(t)$ defined in Eq.~\eqref{eq:charge_t}. Subfigure $(b)$: R\'enyi entropy $S^{(2)}_{N}(t)$ as a function of the charge $\mathcal{Q}_{\mathcal{S}}(t)$ (note the inverted scale on the $x$-axis). The plots correspond to a random evolution in the presence of a global $U(1)$ conserved charge, while the system is initialized as in Eq.~\eqref{eq:initial_state}. The evolution parameters are set to $\lambda_1=1$ and $\lambda_2=2$, and the system size is $N=60$.}
		\label{fig:charge_evolution}
	\end{figure}

	Finally, in order to push further the analogy between our model and a unitary black hole evaporation process, it is interesting to study the time evolution of the system charge 
	\be
	\mathcal{Q}_{\mathcal{S}}(t)={\rm tr}_\mathcal{S}\left\{\mathbb{E}\left[\rho_{\mathcal{S}}(t)\right]\sum_{j\in\mathcal{S}}q_{j}\right\}\,.
	\label{eq:charge_t}
	\ee
	The computation of $\mathcal{Q}_{\mathcal{S}}(t)$ can be carried out using the very same techniques outlined above for the second R\'enyi entropy. In fact, we note that the calculations are simpler, since they only involve two replicas, instead of four. In particular, it turns out that the average charge can be obtained as the solution to a system of $(N+1)^2$ coupled linear differential equations, which can be easily treated numerically. Since no additional complication arises, we omit the details of the computation here, and only report our final numerical results. These are displayed in Fig.~\ref{fig:charge_evolution}, where we also show $S^{(2)}_N(t)$ as a function of $\mathcal{Q}_{\mathcal{S}}(t)$. 
	
	We note that the dimension of the Hilbert space associated with a given integer value $Q$ of the charge is $\binom{N}{Q}$. Of course, the evolved system state will have nonzero projection onto different sectors of the charge at a given time. Nevertheless, we can define an effective ``black hole''  Hilbert space dimension associated with the averaged charge as
		\be
		D_{BH}(t)=\binom{N}{\mathcal{Q}_\mathcal{S}(t)}\,.
		\label{eq:effective_HS}
		\ee
		We see that the behavior of $D_{BH}(t)$ depends on the initial state chosen for $\mathcal{S}$. If the system is initialized as in Eq.~\eqref{eq:initial_state}, then the effective Hilbert space dimension will first increase, and then decrease after $\mathcal{Q}_\mathcal{S}(t)$ reaches the value $N/2$. On the other hand, if the system is initialized in the state~\eqref{eq:initial_state_halfhalf}, the effective Hilbert space decreases monotonically, as one would expect in a more realistic unitary black hole evaporation process.
	
Before leaving this section, we comment on the choice~\eqref{eq:initial_state_environment} for the initial state of the environment. As we have mentioned, this is motivated by an analogy with the black hole evaporation process, where~\eqref{eq:initial_state_environment} plays the role of the ``global'' vacuum state. However, it is natural to wonder what would happen if $\mathcal{E}$ is initialized, instead, in a random product state. In this case, a non-vanishing charge would be pumped into the system at each time interval, so one would expect that the qualitative behavior of the R\'enyi entropy would be similar as in the evolution without the $U(1)$ conserved charge. We have verified by explicit numerical calculations that this is indeed the case. In particular, if  $\mathcal{E}$ is initialized in a random product state we observe no monotonic decrease of the R\'enyi entropy after the Page time.	
	
	\section{Retrieval of quantum information}
	\label{sec:info_retrieval}

	In this section we finally discuss how the RUC introduced in Sec.~\ref{sec:model} provides a microscopic model for the information-retrieval protocol studied by Hayden and Preskill~\cite{hayden2007black}, and allows us to investigate quantitatively several aspects of the latter. We start by briefly reviewing the setting of Ref.~\cite{hayden2007black}, and then proceed to present our results.
	
	We recall that the information stored in a black hole is emitted in the form of Hawking radiation~\cite{Hawking1976}, so that one can ask what is the minimum amount of time that is needed before such information can be recollected from measurements performed outside of the black hole. In order to make contact with our model, we interpret $\mathcal{S}$ as the black hole, while $\mathcal{E}$ consists of all its exterior degrees of freedom (hence including Hawking radiation). Following Ref.~\cite{hayden2007black}, we then imagine that Alice injects a qudit $A$ into the system at time $t=0$, and that a third party $\mathcal{C}$ (Charlie), holds a reference qudit which is maximally entangled with the former. The system is in contact with $\mathcal{E}$ and evolved by the RUC introduced in Sec.~\ref{sec:model}. Finally, we imagine that Bob wants to recover information on the injected qudit by only performing measurements outside of $\mathcal{S}$. Depending on the initial configuration of the system, the ability to faithfully do so after a given time $t$ is captured quantitatively by the mutual information between different sets of qudits, as we  now explain.
	
	\begin{figure}
		\begin{tabular}{ll}
			\hspace{-0.25cm}\includegraphics[width=0.35\textwidth]{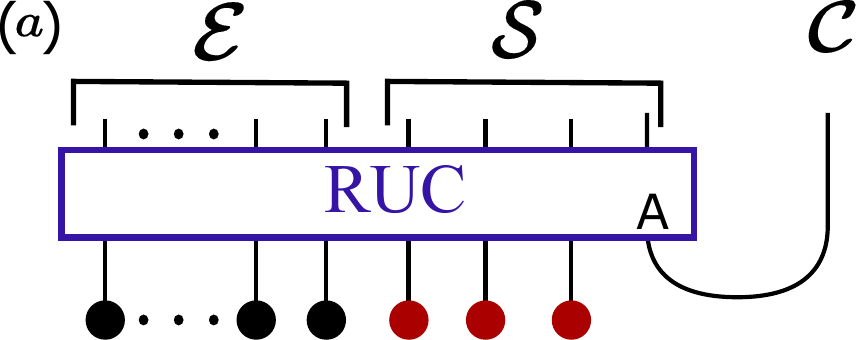} & \hspace{2.cm}\includegraphics[width=0.42\textwidth]{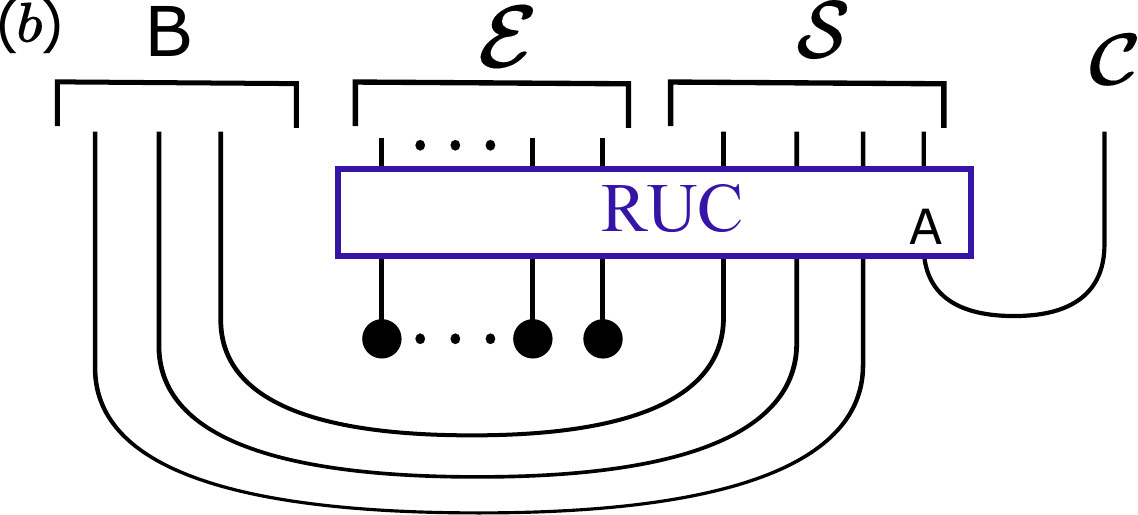}
		\end{tabular}
		\caption{Pictorial representation of the two settings considered in Sec.~\ref{sec:info_retrieval}. In the first case [subfigure $(a)$], we initialize the system $\mathcal{S}$ (the ``black hole'') in a product state. A qudit $A$ is initially injected into the black hole, and a third party ($\mathcal{C}$) holds a reference system, namely an ancilla which is maximally entangled with the former at time $t=0$. The system is in contact with $\mathcal{E}$ (representing the exterior of the black hole, including Hawking radiation) and evolved by the RUC introduced in Sec.~\ref{sec:model}. In the second setting [subfigure $(b)$], the retriever ($B$) initially holds a copy of the black hole, namely $\mathcal{S}$ is initialized in a maximally entangled state with a set of ancillary qudits.}
		\label{fig:models_initial confuguration}
	\end{figure}
	
	First, let us consider the setting pictorially depicted in Fig.~\ref{fig:models_initial confuguration}$(a)$: in this case, Bob has no control over the initial configuration of the system $\mathcal{S}$, which is initialized in a given product state at time $t=0$. The capability of recovering information on the injected qudit by measurements on $\mathcal{E}$ is quantified by the mutual information
	\be
	I_{(a),[\mathcal{C},\mathcal{E}]}(t)={S}_\mathcal{C}(t)+{S}_\mathcal{E}(t)-{S}_{\mathcal{C}\cup \mathcal{E}}(t)\,,
	\label{eq:neumann_mutual_info}
	\ee
	which tells us how much information can be extracted from the reference qudit $\mathcal{C}$ by accessing those in $\mathcal{E}$. In particular, if $I_{(a),[\mathcal{C},\mathcal{E}]}(t)$ is close to its maximal value, then Bob can faithfully recover the information initially injected into $\mathcal{S}$. Note that in Eq.~\eqref{eq:neumann_mutual_info} we used an index $(a)$ to distinguish the two settings in Fig.~\ref{fig:models_initial confuguration}.  As usual, due to computational limitations, in the following we will not compute the quantity in Eq.~\eqref{eq:neumann_mutual_info}, but rather its R\'enyi-$2$ version, namely
	\be
	I^{(2)}_{(a),\left[\mathcal{C},\mathcal{E}\right]}(t)={S}^{(2)}_\mathcal{C}(t)+{S}^{(2)}_\mathcal{E}(t)-{S}^{(2)}_{\mathcal{C}\cup \mathcal{E}}(t)\,,
	\label{eq:renyi_bob_doesnt_know}
	\ee
	where $\mathcal{S}^{(2)}_K(t) $, for a given system $K$, is defined in Eq.~\eqref{eq:log_purity}.
	
	In the second setting, displayed in Fig.~\ref{fig:models_initial confuguration}$(b)$, we imagine instead that the black hole formed long ago, and that Bob has been collecting its emitted Hawking radiation ever since. Accordingly, by the time the qudit $A$ is injected, the black hole $\mathcal{S}$ is in  a maximally entangled state with the previously emitted radiation system, which is under Bob's control [region $B$ in Fig.~\ref{fig:models_initial confuguration}$(b)$]. In this case, Bob can also access these qudits, together with those in the environment $\mathcal{\mathcal{E}}$, and his capability to recover the initially injected information is quantified by $I_{(b),\left[\mathcal{C},\mathcal{E}\cup B\right]}(t)$. Accordingly, analogously to the previous case, in the following we will compute its R\'enyi-$2$ counterpart
	\be
	I^{(2)}_{(b),\left[\mathcal{C},\mathcal{E}\cup B\right]}(t)={S}^{(2)}_\mathcal{C}(t)+{S}^{(2)}_{\mathcal{E}\cup B}(t)-{S}^{(2)}_{\mathcal{C}\cup \mathcal{E}\cup B}(t)\,.
	\label{eq:renyi_bob_knows}
	\ee

	It turns out that the formalism introduced in the previous section is adequate to compute numerically the mutual information in Eqs.~\eqref{eq:renyi_bob_doesnt_know} and \eqref{eq:renyi_bob_knows}, for both RUCs without and with a conserved $U(1)$ charge. To see this, we can exploit the fact that the R\'enyi entropy of a subsystem $K$ is equal to that of its complementary one (with respect to the whole space), and rewrite
	\bea
	I^{(2)}_{(a),\left[\mathcal{C},\mathcal{E}\right]}(t)&=&{S}^{(2)}_{\mathcal{C}}(t)+{S}^{(2)}_{\mathcal{S}\cup \mathcal{C}}(t)-{S}^{(2)}_{\mathcal{S}}(t)\,,
	\label{eq:renyi_bob_doesnt_know_II}\\
	I^{(2)}_{(b),\left[\mathcal{C},\mathcal{E}\cup B\right]}(t)&=&{S}^{(2)}_{\mathcal{C}}(t)+{S}^{(2)}_{\mathcal{S}\cup \mathcal{C}}(t)-{S}^{(2)}_{\mathcal{S}}(t)\,.
	\label{eq:renyi_bob_knows_II}
	\eea
	Each individual entropy in the r.h.s. of the above equations can be computed by exploiting the approach in Sec.~\ref{sec:entropy_no_cons} (for the maximally chaotic RUC) and in Sec.~\ref{sec:entropy_u1} (in the presence of a conserved $U(1)$ charge). In particular, in each case we can map the problem onto the computation of the time evolution  in a four replica space $\widetilde{H}_{\mathcal{S}}$, where the dynamics is driven by the Lindbladian operator \eqref{eq:lindbladian_final}. The only difference with respect to the steps presented in the previous section is in the initial state and purity vector $\langle\bra{W_K}$, which have to be modified for each individual term in the r.h.s of Eqs.~\eqref{eq:renyi_bob_doesnt_know_II} and \eqref{eq:renyi_bob_knows_II}.  Since these calculations do not present additional difficulties,  we report them in Appendix~\ref{sec:details_mutual_info}, and in the rest of this section we present our final results.
	
	\begin{figure}
		\begin{tabular}{ll}
			\hspace{-0.25cm}\includegraphics[width=0.48\textwidth]{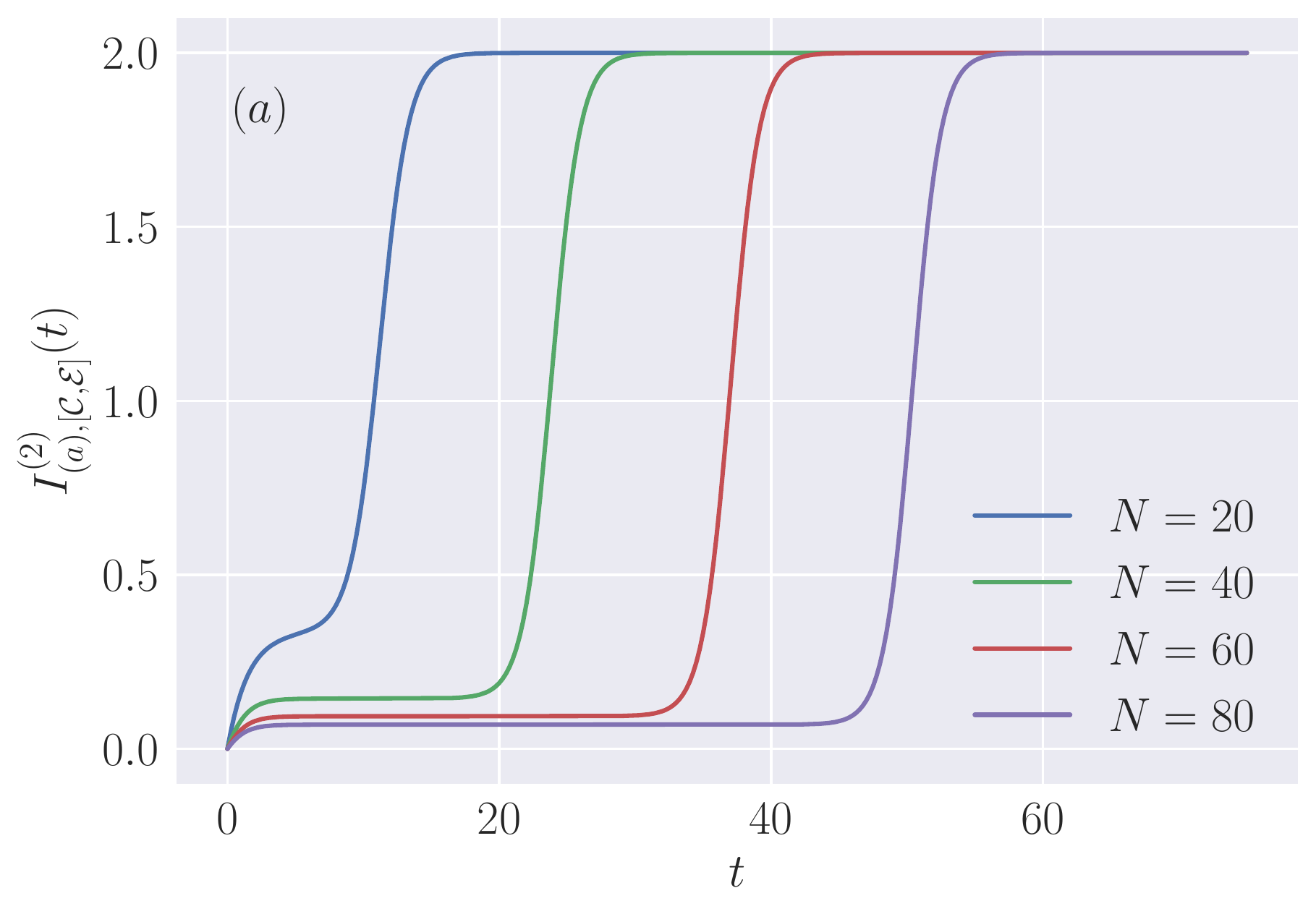} & \hspace{0.cm}\includegraphics[width=0.48\textwidth]{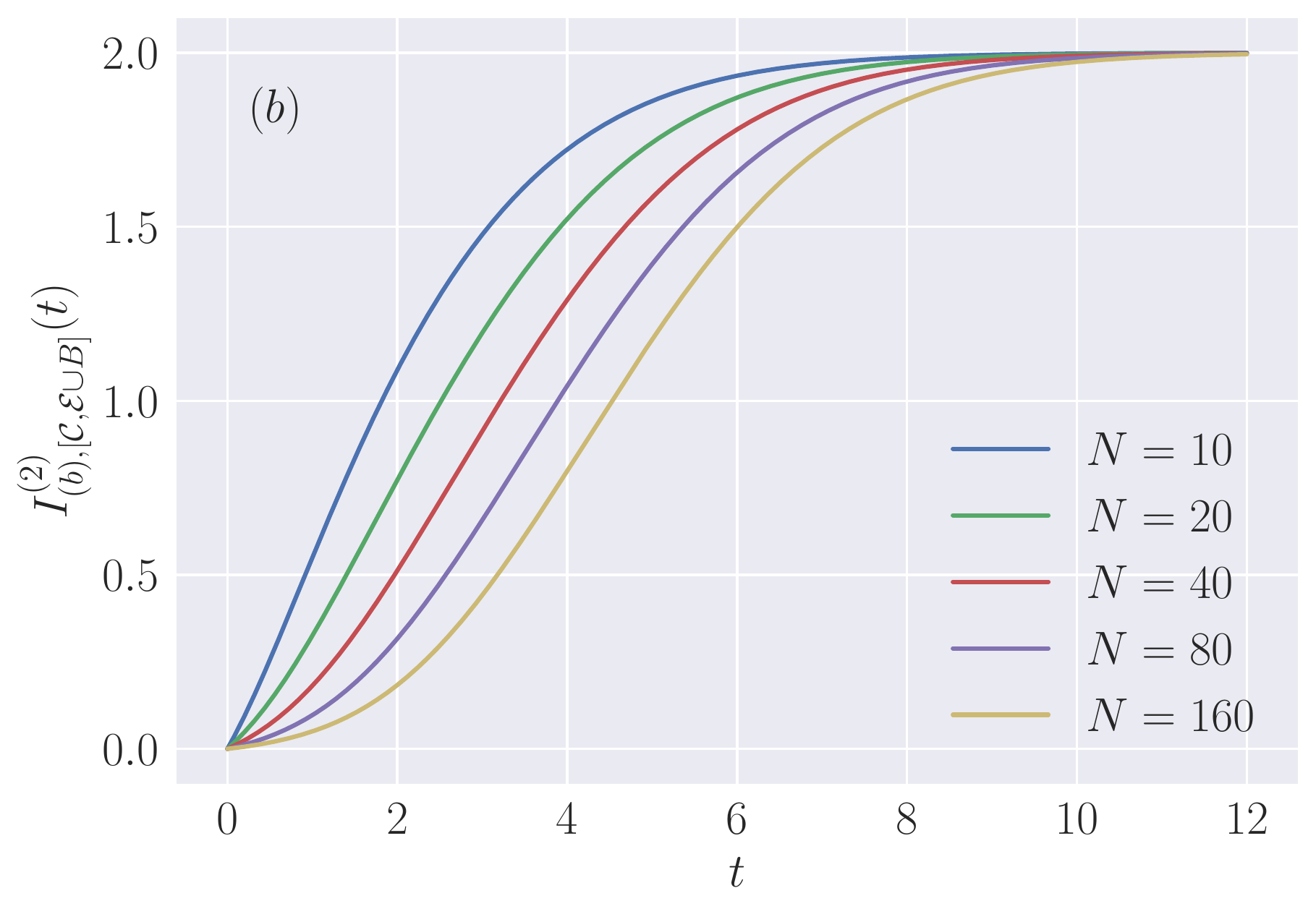}
		\end{tabular}
		\caption{Subfigure $(a)$: mutual information $I^{(2)}_{(a),\left[\mathcal{C},\mathcal{E}\right]}(t)$ [defined in Eq.~\eqref{eq:renyi_bob_doesnt_know}] for the setting displayed in  Fig.~\ref{fig:models_initial confuguration}$(a)$, and increasing values of $N$. Subfigure $(b)$: mutual information $I^{(2)}_{(b),\left[\mathcal{C},\mathcal{E}\cup\mathcal{B}\right]}(t)$ [defined in Eq.~\eqref{eq:renyi_bob_knows}] for the setting displayed in  Fig.~\ref{fig:models_initial confuguration}$(b)$, and increasing values of $N$. For both plots, the evolution is driven by the maximally chaotic RUC of Sec.~\ref{sec:model} (without conserved charges), where we set $\lambda_1=1$, $\lambda_2=2$ and chose $d=2$.}
		\label{fig:mutual_info_unconserved}
	\end{figure}
	
	\begin{figure}
		\begin{tabular}{ll}
			\hspace{-0.25cm}\includegraphics[width=0.48\textwidth]{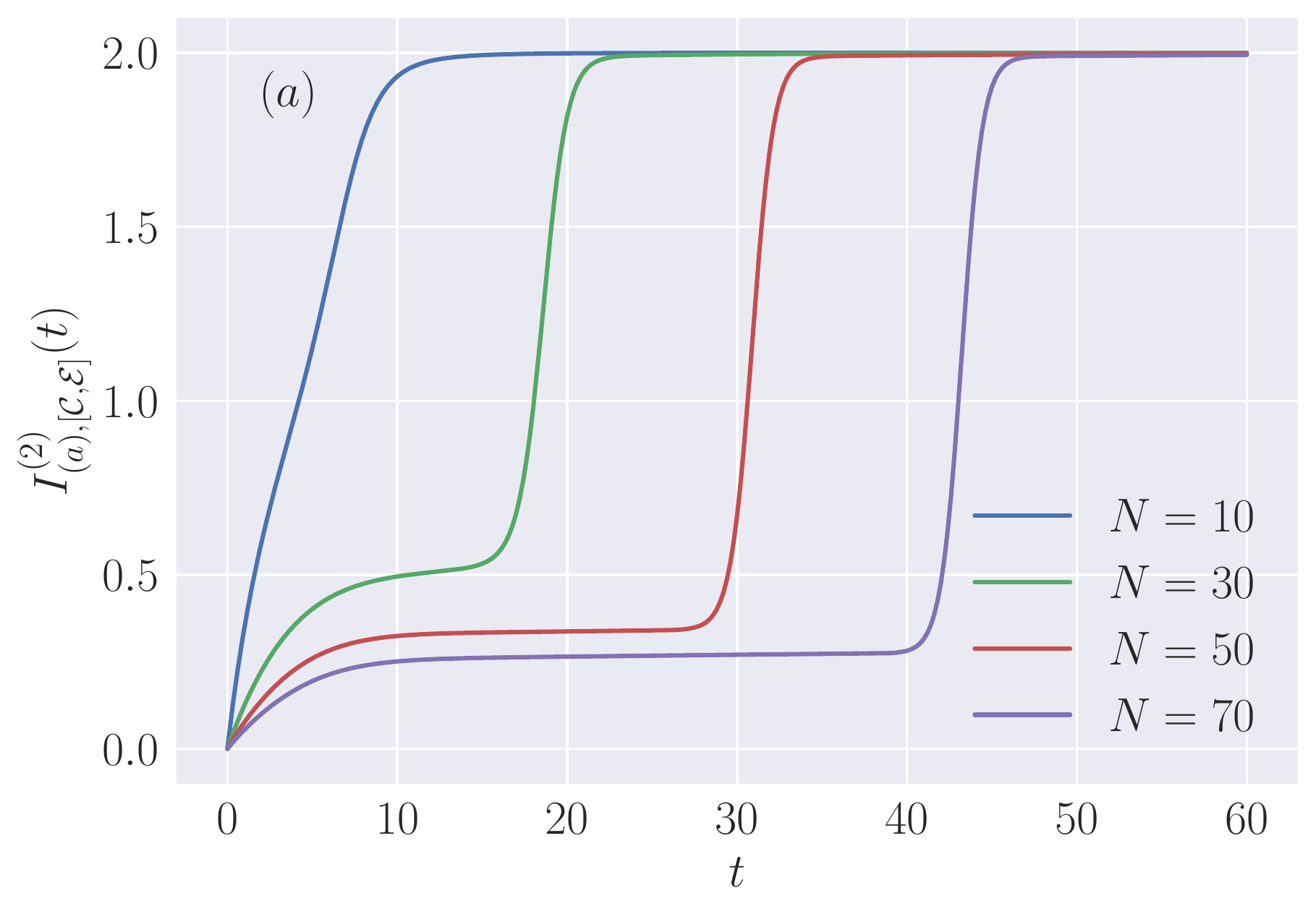} & \hspace{0.cm}\includegraphics[width=0.48\textwidth]{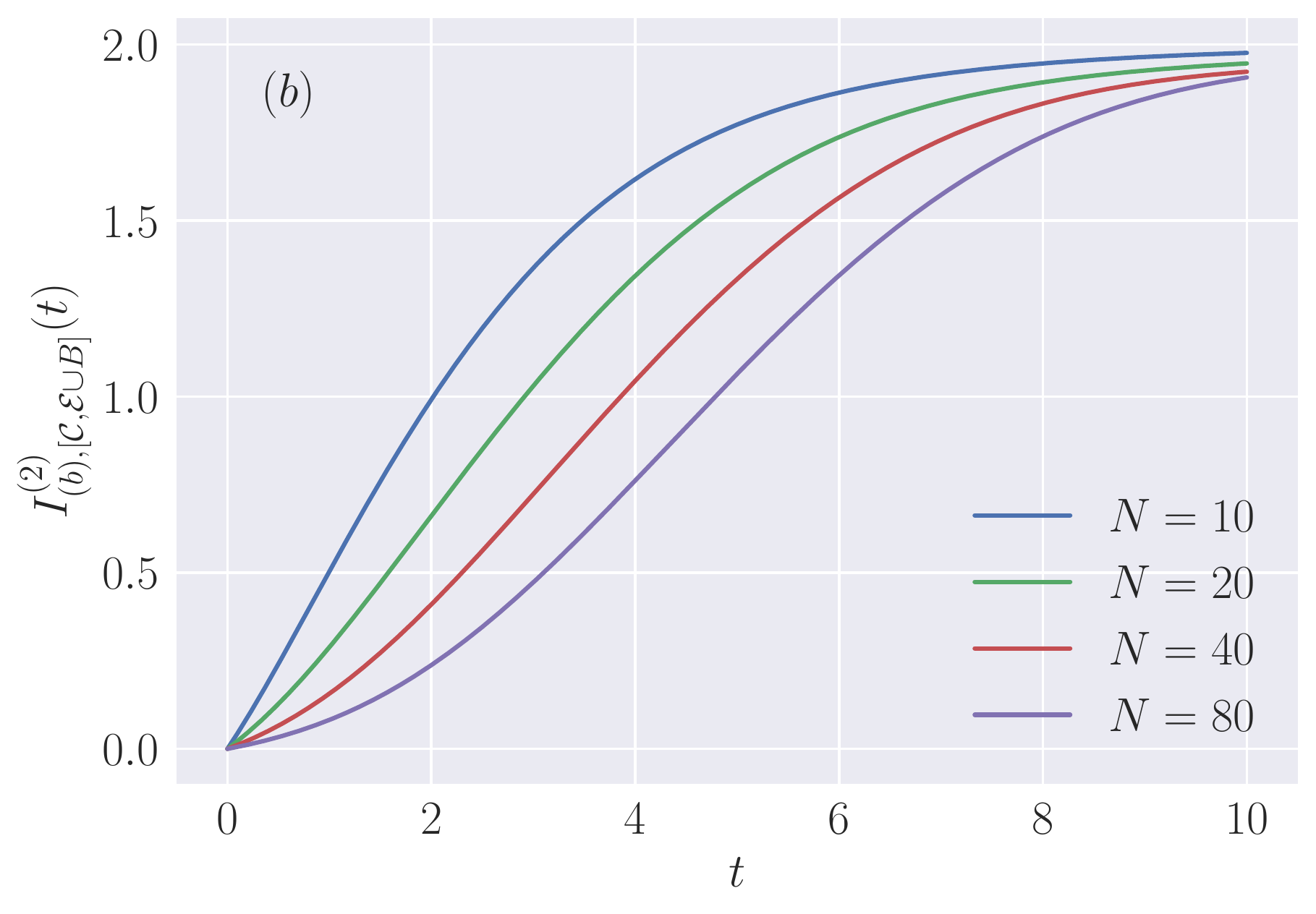}
		\end{tabular}
		\caption{Same plots as in Fig.~\ref{fig:mutual_info_unconserved}. The evolution is now driven by the RUC with a conserved $U(1)$ charge defined in Sec.~\ref{sec:entropy_u1}, where we set $\lambda_1=1$, $\lambda_2=2$ (and $d=2$).}
		\label{fig:mutual_info_conserved}
	\end{figure}
	
	We begin by discussing Fig.~\ref{fig:mutual_info_unconserved}, where we report data for the maximally chaotic RUC (no conserved charge). Subfigures $(a)$ and $(b)$ correspond to the two different settings discussed above, and display respectively $I^{(2)}_{(a),\left[\mathcal{C},\mathcal{E}\right]}(t)$ and $I^{(2)}_{(b),\left[\mathcal{C},\mathcal{E}\cup B\right]}(t)$  for increasing values of  $N$. In both cases,  the mutual information has a monotonic behavior, although with qualitative  differences. In the first case, it reaches its maximum value in a time which is clearly proportional to the system size $N$.  Interestingly, we see that  after a time scale of the order of the scrambling  time $t_s\sim\ln N$, the mutual information reaches a small non-zero value, which, however, is seen to  decrease with the system  size $N$. We can interpret this as follows: after the scrambling time, Bob is able to only reconstruct a small amount of the initially injected information, and needs to wait for a time proportional to the black hole size in order to retrieve all of it. 
	
	Conversely, we see from Fig.~\ref{fig:mutual_info_unconserved}$(b)$ that the information retrieval is much faster in the case Bob holds a copy of the black hole [cf. Fig.~\ref{fig:models_initial confuguration}$(b)$]. In particular, from the plot we clearly see that the mutual information reaches its maximum value after a time which is logarithmic with the system size $N$, namely  the scrambling time. 
	
	We have repeated the same calculations for a RUC with a conserved $U(1)$ charge, and reported our results in Fig.~\ref{fig:mutual_info_conserved}. We see that the functions $I^{(2)}_{(a),\left[\mathcal{C},\mathcal{E}\right]}(t)$ and $I^{(2)}_{(b),\left[\mathcal{C},\mathcal{E}\cup B\right]}(t)$ display the same qualitative features. It is interesting to note that, in the setting corresponding to subfigure $(a)$ of Fig.~\ref{fig:models_initial confuguration}, the value of the mutual information after the  scrambling  time is larger than that in the maximally chaotic case, although still vanishing for $N\to  \infty$. This is intuitive: the presence of conservation laws constrains the Hilbert space that can be explored by the system, hence generally increasing the knowledge on its state. 
	
		\begin{figure}
		\begin{tabular}{lll}
			\hspace{-0.35cm}\includegraphics[width=0.35\textwidth]{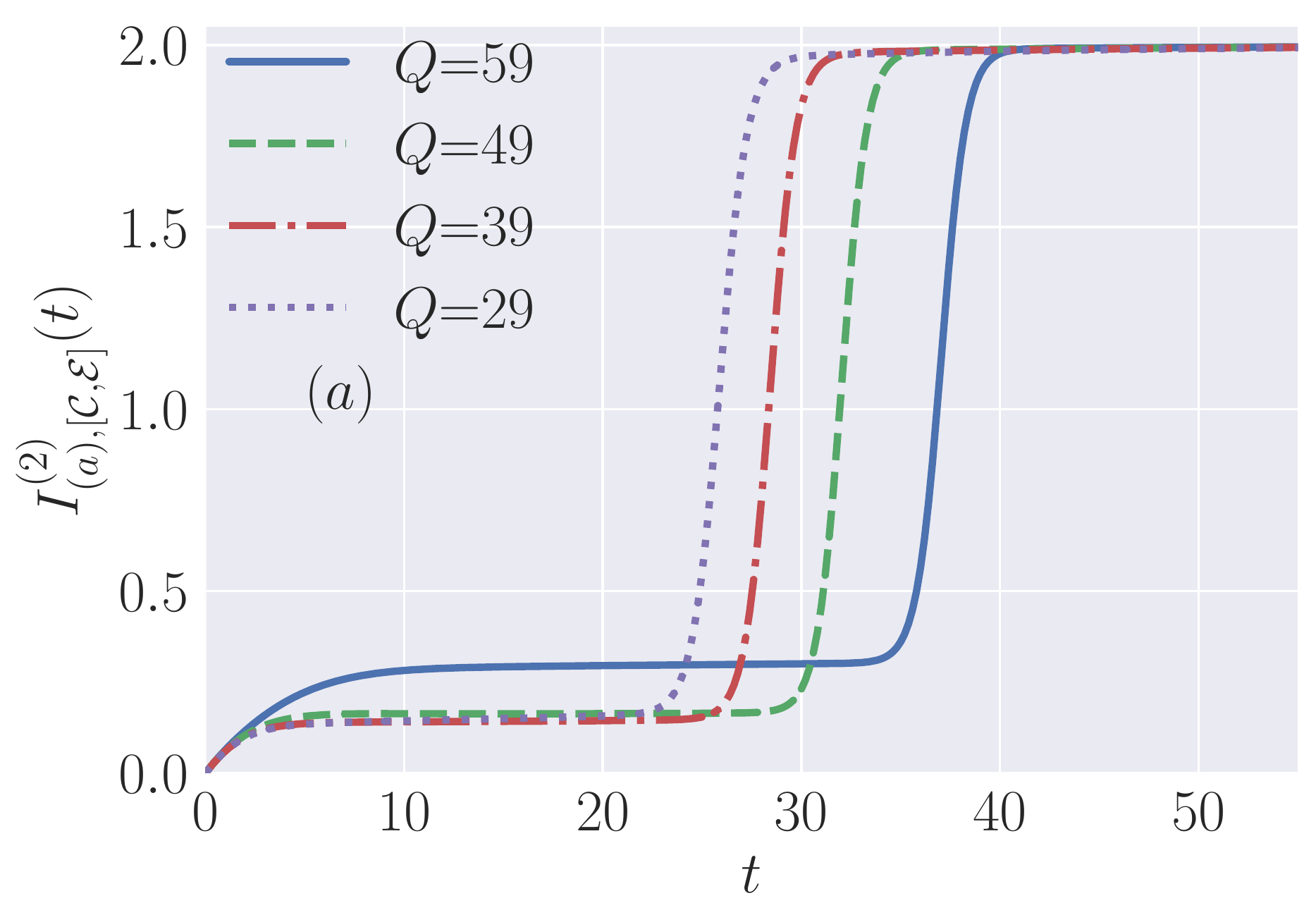} & \hspace{-0.35cm}\includegraphics[width=0.35\textwidth]{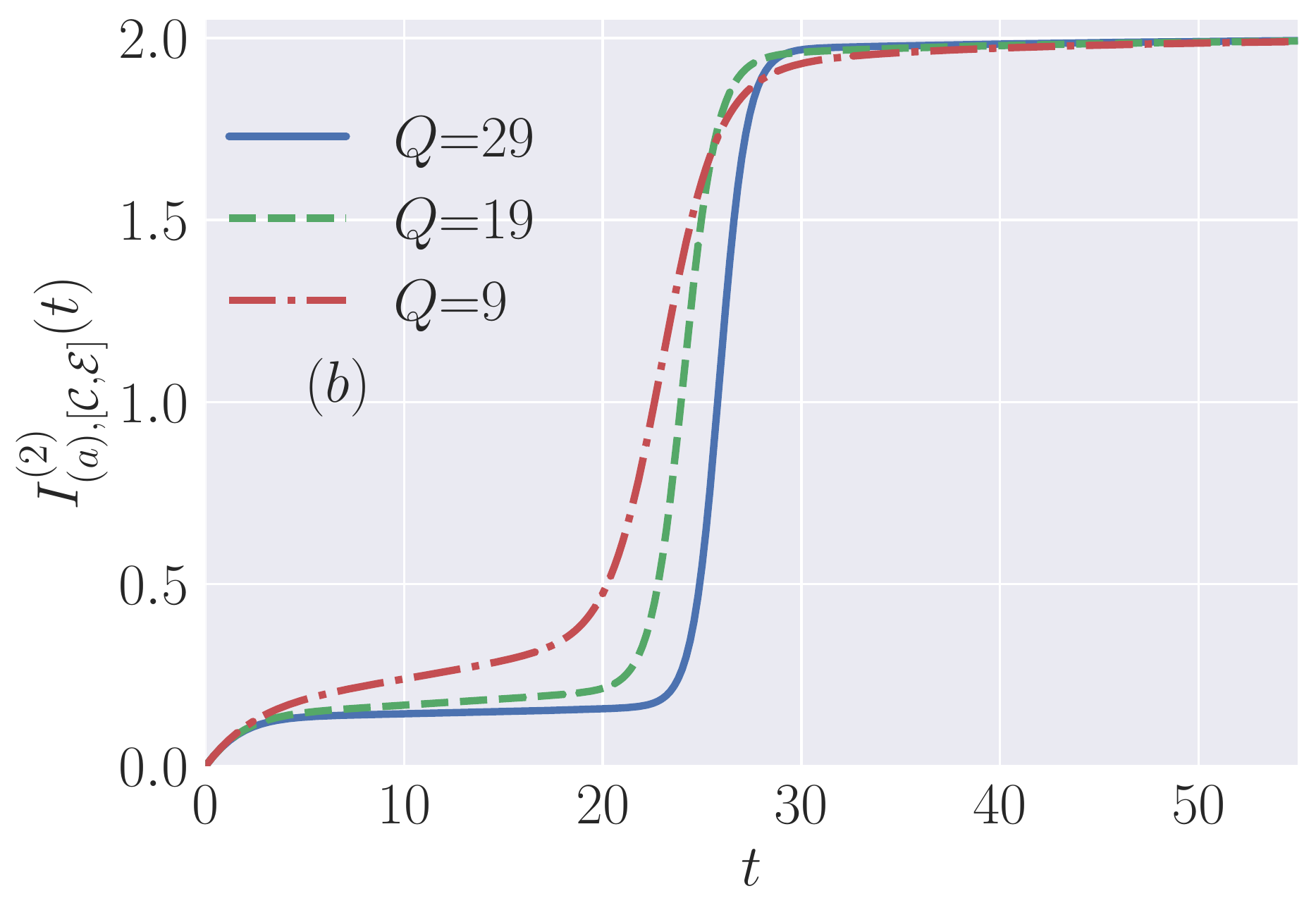}& \hspace{-0.35cm}\includegraphics[width=0.35\textwidth]{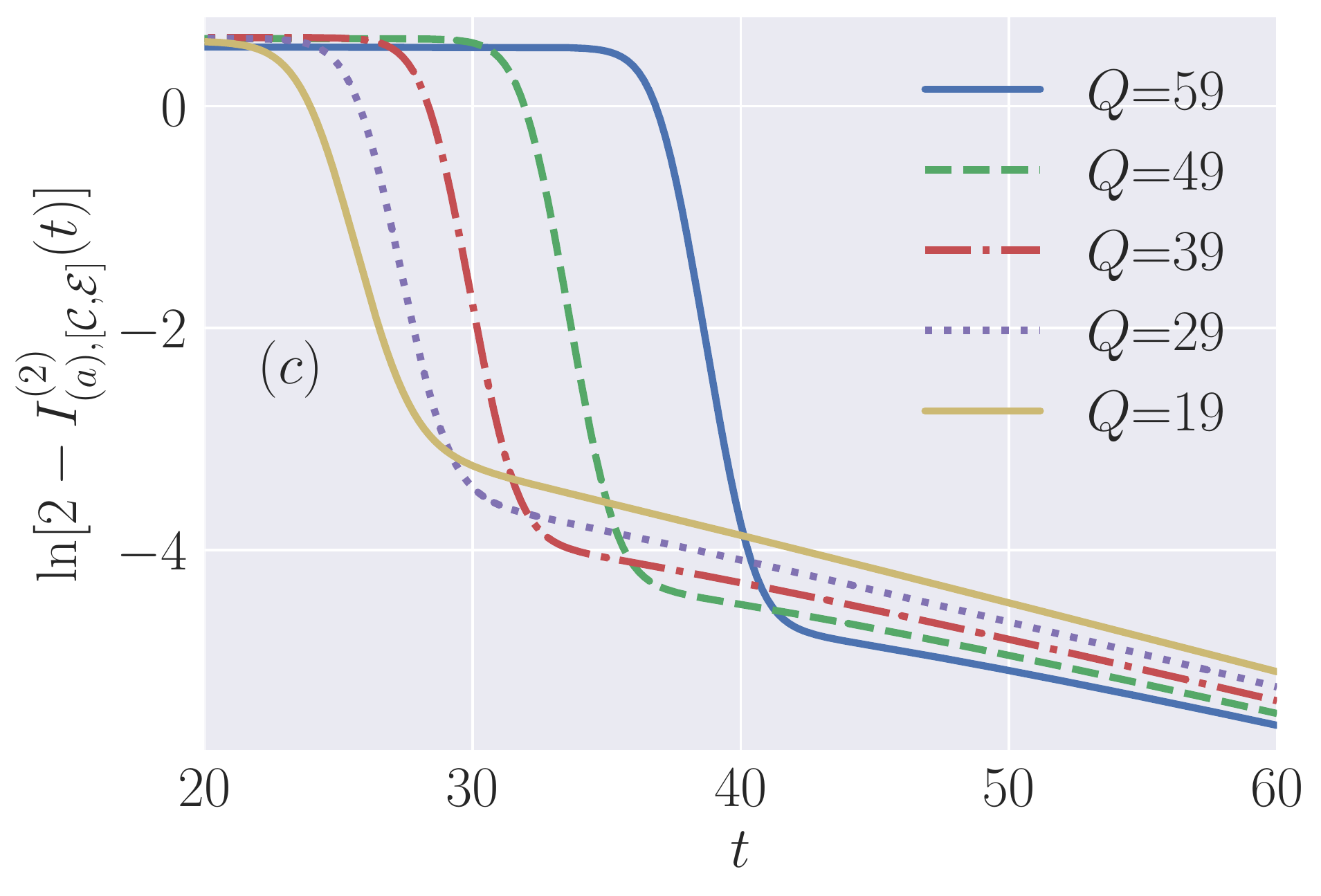}
		\end{tabular}
		\caption{Mutual information $I^{(2)}_{(a),\left[\mathcal{C},\mathcal{E}\right]}(t)$ [defined in Eq.~\eqref{eq:renyi_bob_doesnt_know}] for the setting displayed in  Fig.~\ref{fig:models_initial confuguration}$(a)$.  Subfigures $(a)$ and $(b)$ correspond to the initial state~\eqref{eq:initial_state_n_dep} for different values of the initial charge $Q=N-1-n$, respectively larger and smaller than $N/2$. Subfigure $(c)$ shows $\ln \left[2-I^{(2)}_{(a),\left[\mathcal{C},\mathcal{E}\right]}(t)\right]$ for the same state (and different values of $Q$) at late times.  For all plots, the evolution is driven by the RUC  with a conserved $U(1)$ charge defined in Sec.~\ref{sec:entropy_u1}, where we set $\lambda_1=1$, $\lambda_2=2$ and $N=60$ (while we chose $d=2$).}
		\label{fig:mutual_info_conserved_different_states}
	\end{figure}

In energy conserved systems, the Lyapunov exponent measured by OTOC growth generically depends on temperature. Usually a slower scrambling occurs at lower temperature $T$, and an upper bound of $2\pi T$ has been proven for a particular regularized version of the OTOC \cite{maldacena2016bound}. The analog of temperature dependence in our model is the charge dependence of the information retrieval time. We expect that when the charge is closer to $0$ or $1$, the Hilbert space size is smaller, leading to a similar effect as reducing temperature in energy conserved system. For this purpose, we study the mutual information growth for states with different charge. To this end, we first consider the protocol depicted in Fig.~\ref{fig:models_initial confuguration}$(a)$, but we initialize the system $\mathcal{S}$ in the product state
		\be
		\ket{\Psi^{\mathcal{S}}_0}=\bigotimes_{j=1}^{n}\ket{0}_j \bigotimes_{j=n+1}^{N-1} \ket{1}_j\,. 
		\label{eq:initial_state_n_dep}
		\ee
		As we have clarified after Eq.~\eqref{eq:initial_state_halfhalf}, we actually consider averages over all the initial states obtained by permuting different qubits in Eq.~\eqref{eq:initial_state_n_dep}, namely over all the product states with $n$ qubits initialized to $\ket{0}$ and $(N-1-n)$ to $\ket{1}$ (where the last qubit $N$ corresponds to $A$, and is entangled to the ancilla $\mathcal{C}$). This allows us  to exploit the permutational symmetry in the four-replica space, and proceed following the very same steps outlined above to obtain numerically exact data for the mutual information $I^{(2)}_{(a),\left[\mathcal{C},\mathcal{E}\right]}(t)$. 
	
 We report our results in Fig.~\ref{fig:mutual_info_conserved_different_states}, for different values of $n$, and a fixed system size $N=60$. In subfigures $(a)$ and $(b)$ we report data for decreasing values of the initial charge $Q=N-1-n$, respectively larger or smaller than $N/2$. As we have already pointed out, in the former case the effective Hilbert space dimension~\eqref{eq:effective_HS} has a non-monotonic behavior, whereas in the latter case it is monotonically decreasing, as one would expect in a more realistic unitary evaporation protocol. This is  reflected in the fact that, at short times, the two plots display a different qualitative behavior as $n$ increases: in subfigure $(a)$, $I^{(2)}_{(a),\left[\mathcal{C},\mathcal{E}\right]}(t)$ decreases as $n$ increases, while the opposite happens in subfigure~$(b)$. 
	
In subfigure $(c)$ we report instead the logarithm of the difference between the maximum value $2$ of the mutual information and $I^{(2)}_{(a),\left[\mathcal{C},\mathcal{E}\right]}(t)$ at late times. The plot shows the emergence of an exponential decay, which starts first for smaller initial charge (a larger initial charge takes longer to evaporate).

\begin{figure}
	\includegraphics[scale=0.7]{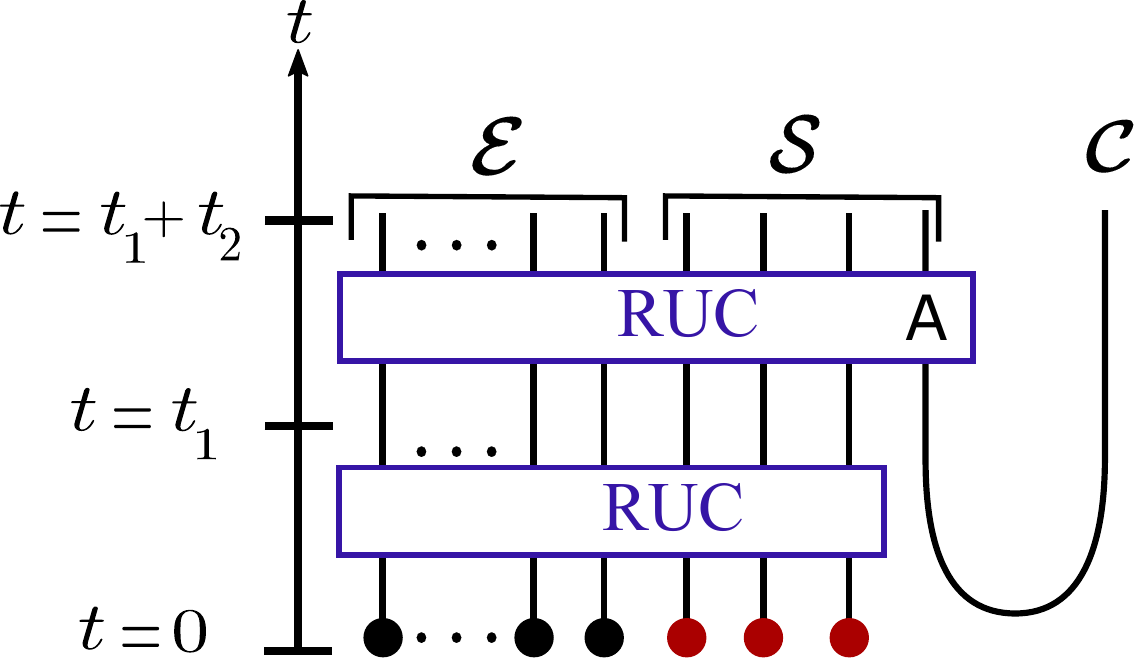}
	\caption{Pictorial representation of the third settings considered in Sec.~\ref{sec:info_retrieval}. We initialize the system $\mathcal{S}$ (the ``black hole'') in a product state, except for a qubit $A$, randomly chosen, which is maximally entangled with an ancillary one, denoted by $\mathcal{C}$. The system $\mathcal{E}\cup \mathcal{S}\setminus A$ is evolved with the RUC with a $U(1)$ conserved charge for a time $t_1$. After that, $\mathcal{E}\cup \mathcal{S}$ is evolved with the same RUC, for a time $t_2$.}
	\label{fig:model_later_times}
\end{figure}

\begin{figure}
	\begin{tabular}{ll}
		\hspace{-0.25cm}\includegraphics[width=0.48\textwidth]{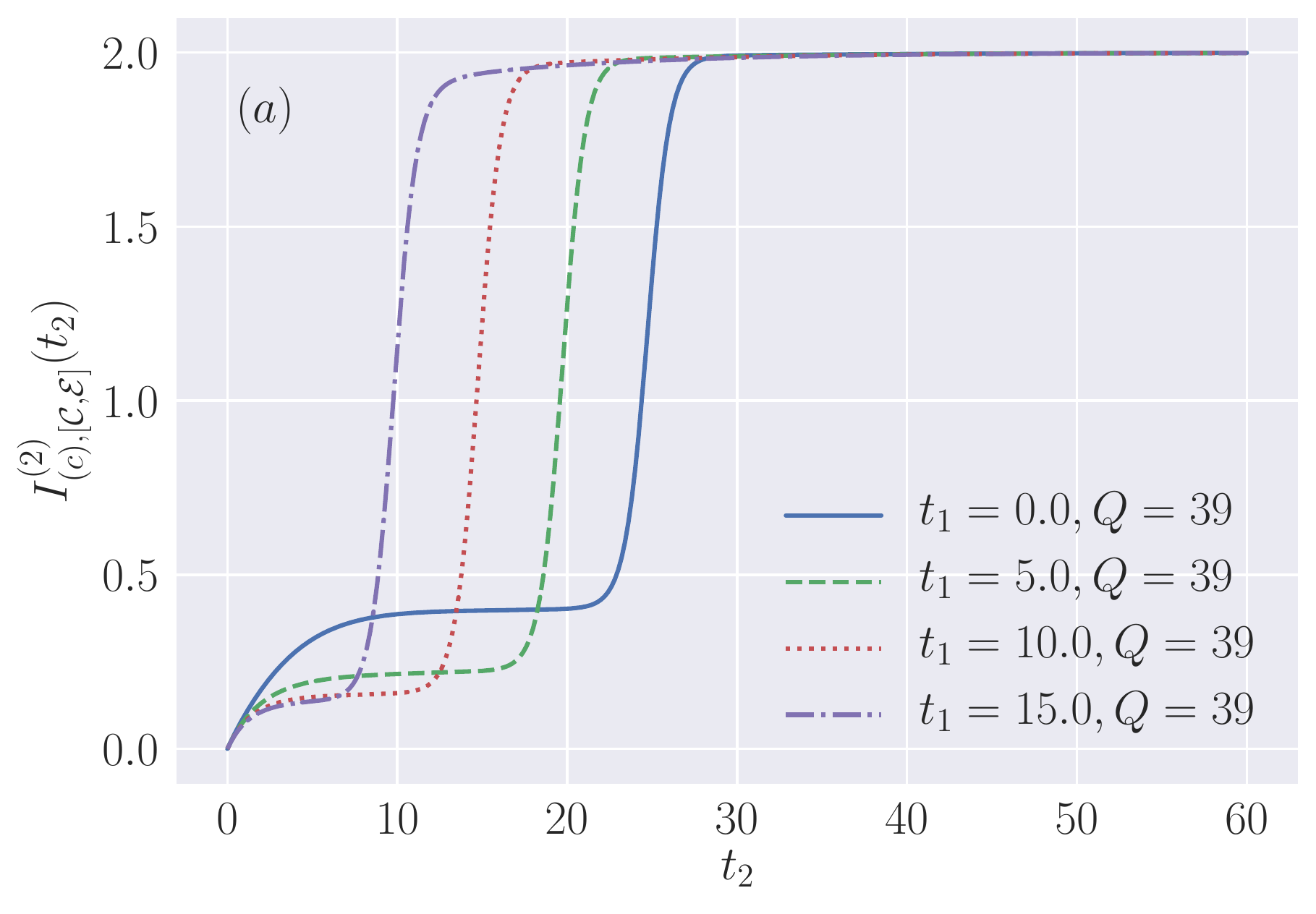} & \hspace{0cm}\includegraphics[width=0.48\textwidth]{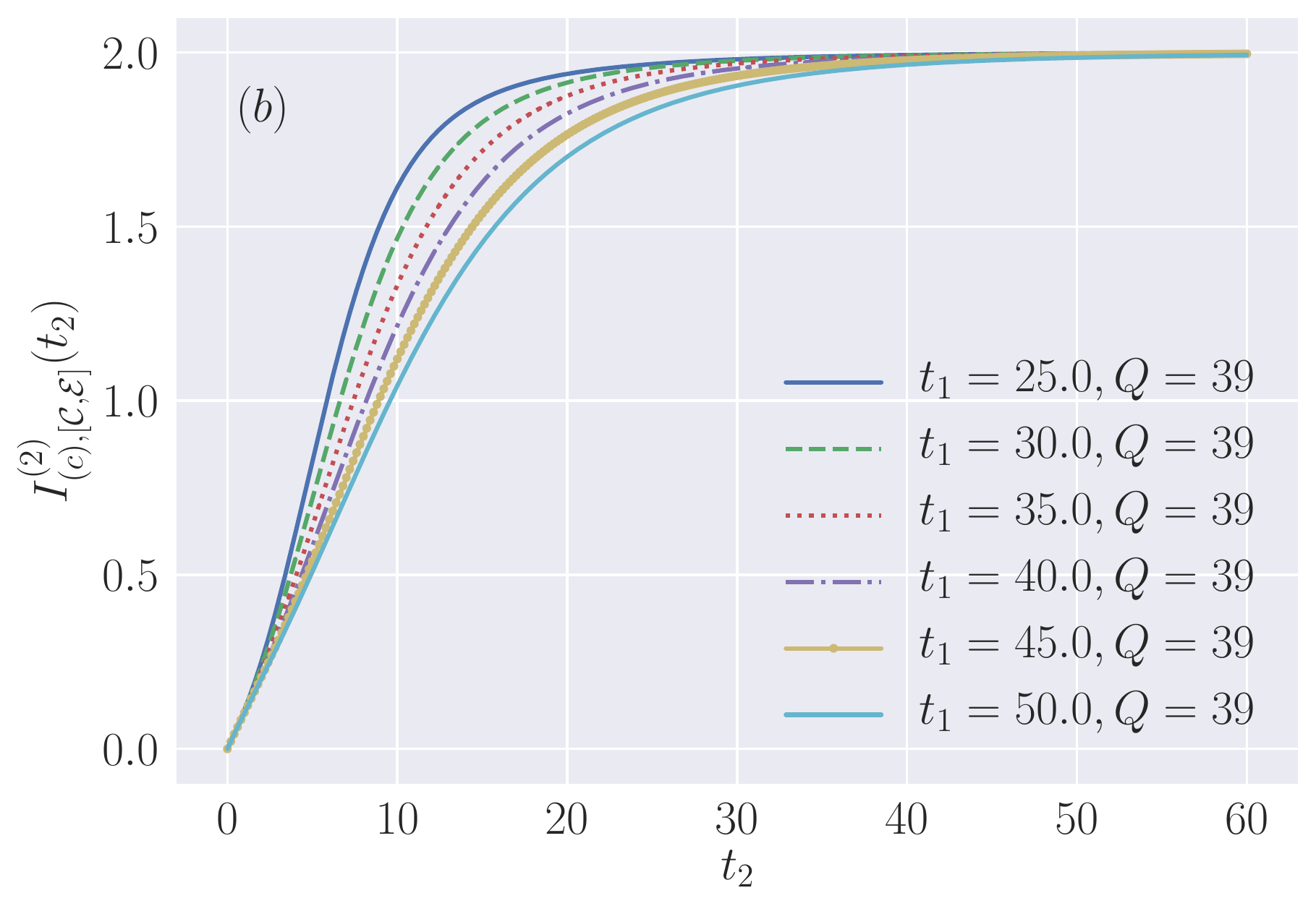}
	\end{tabular}
	\caption{Mutual information $I^{(2)}_{(c),\left[\mathcal{C},\mathcal{E}\right]}(t_2)$ [defined in Eq.~\eqref{eq:mutual_renyi_c}] for the setting displayed in  Fig.~\ref{fig:model_later_times}. Subfigures $(a)$ and $(b)$ show the mutual information for initial charge $Q=39$ and increasing values of $t_1$, respectively smaller and larger than the Page time $t_p$. For all plots, the evolution is driven by the RUC with a conserved $U(1)$ charge defined in Sec.~\ref{sec:entropy_u1}, where we set $\lambda_1=1$, $\lambda_2=2$ and $N=40$ (while we chose $d=2$). }
	\label{fig:mutual_info_conserved_later_times}
\end{figure}
\begin{figure}
	\begin{tabular}{ll}
		\hspace{-0.25cm}\includegraphics[width=0.47\textwidth]{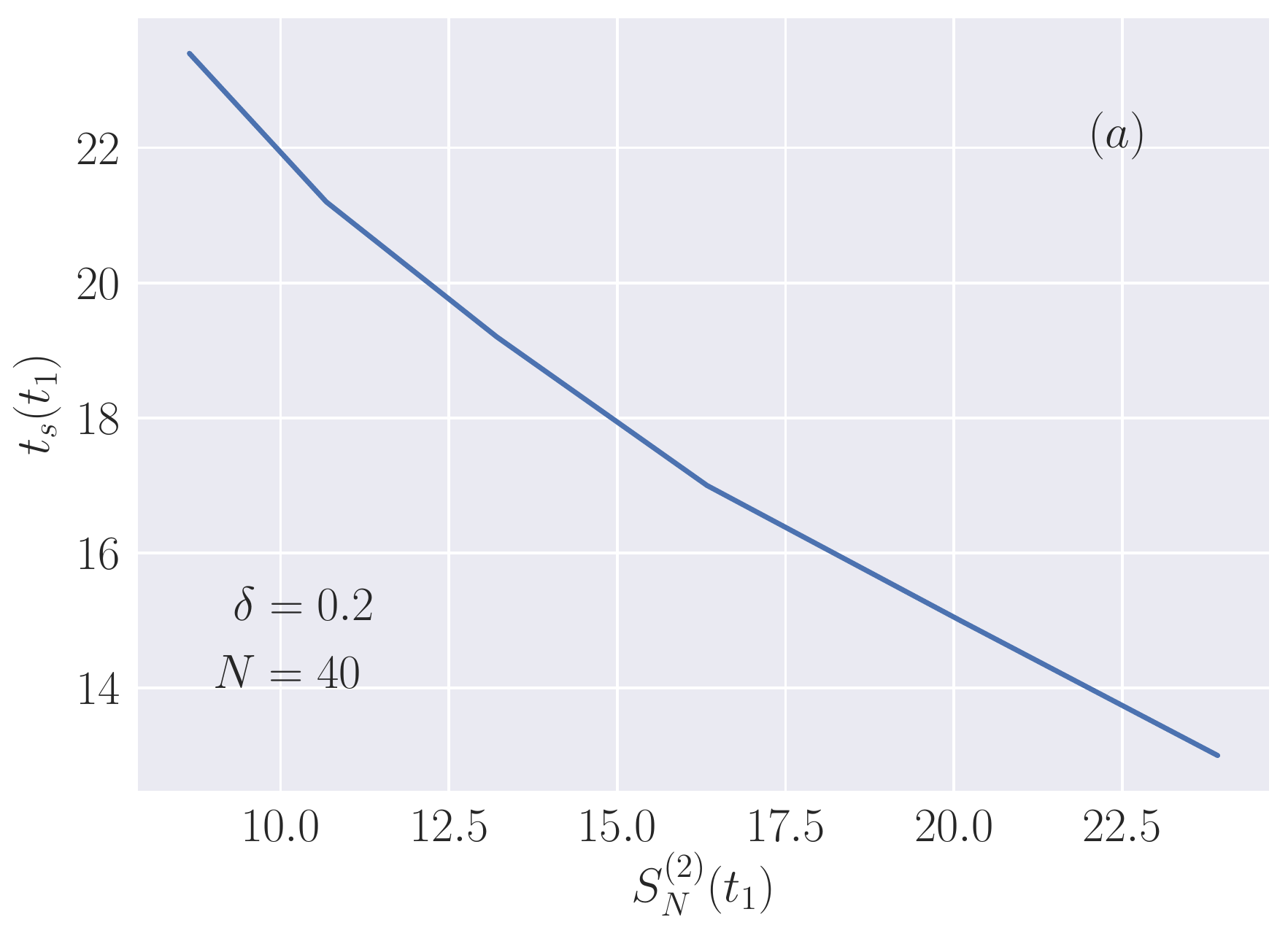} & \hspace{0cm}\includegraphics[width=0.485\textwidth]{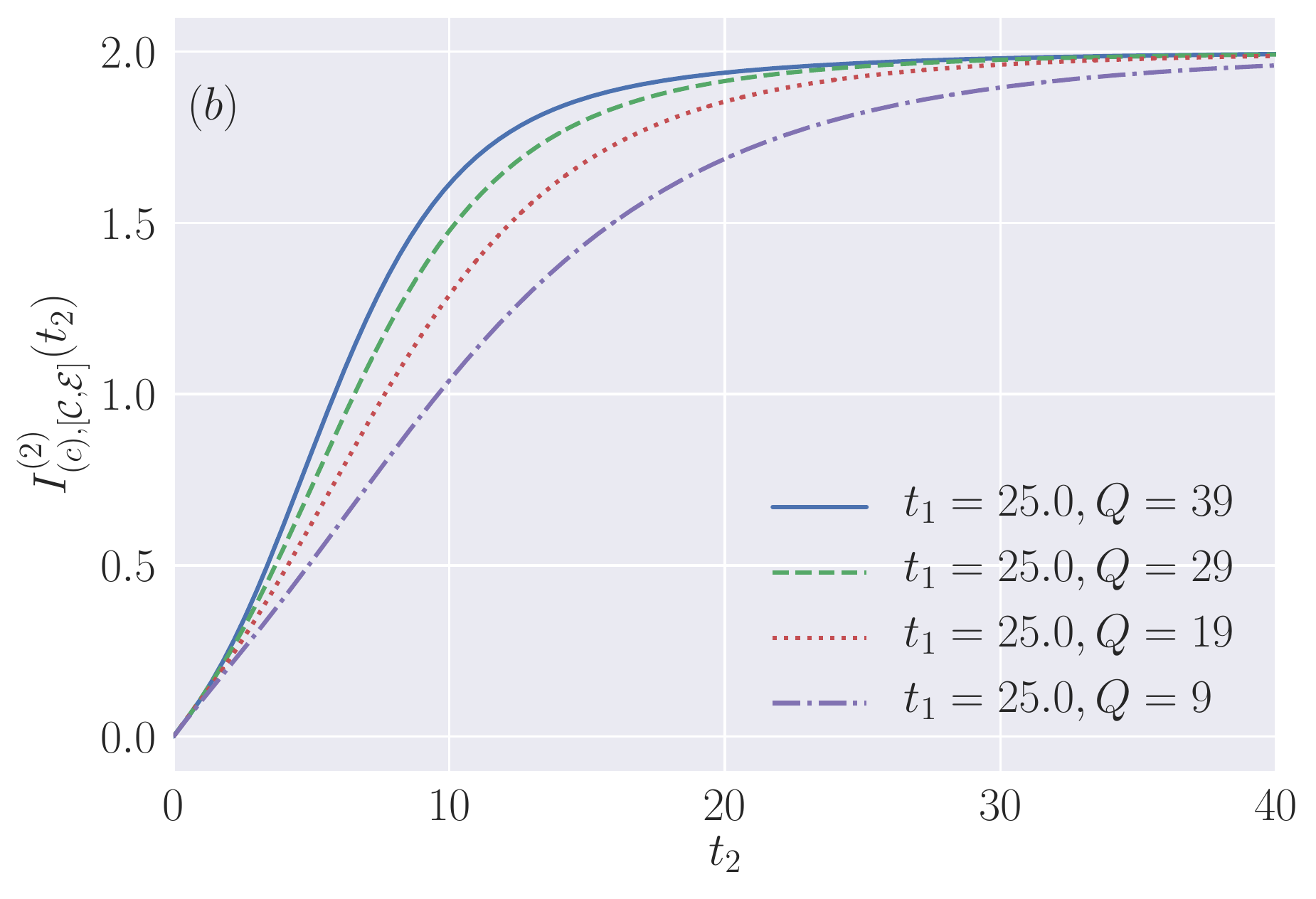}
	\end{tabular}
	\caption{Subfigure $(a)$: scrambling time $t_s(t_1)$ as a function of $S^{(2)}_N(t_1)$ . Here $t_s(t_1)$ is defined as the value of $t_2$ at which $I^{(2)}_{(c),\left[\mathcal{C},\mathcal{E}\right]}(t_2)$  reaches the value $2-\delta$. In the plot, we chose $\delta=0.2$. Subfigure $(b)$: mutual information $I^{(2)}_{(c),\left[\mathcal{C},\mathcal{E}\right]}(t_2)$ for various initial charges $Q$ and fixed $t_1$. For all plots, the evolution is driven by the RUC  with a conserved $U(1)$ charge defined in Sec.~\ref{sec:entropy_u1}, where we set $\lambda_1=1$, $\lambda_2=2$ and $N=40$ (while we chose $d=2$). }
	\label{fig:mutual_info_conserved_later_times_II}
\end{figure}

Next, we consider the protocol reported in Fig.~\ref{fig:models_initial confuguration}$(b)$. In this case, the initial state is given by a maximally entangled state between $\mathcal{S}$ and $B$. This has a non-vanishing projection over all the charge sectors, so we can not vary arbitrarily its charge, as for Fig.~\ref{fig:models_initial confuguration}$(a)$. For this reason, we consider a different setting, which maintains some of its features, but allows us to tune the initial charge of $\mathcal{S}$. This is depicted in Fig.~\ref{fig:model_later_times}. The idea is to initialize the system in a product state, and let the RUC generate an entangled state between $\mathcal{S}$ and $\mathcal{E}$. After the Page time, $\mathcal{S}$ is approximately a maximally mixed state in a certain charge sector (with charge decreasing in time), as we discussed earlier. At time $t=t_1$, we introduce a new qubit which is maximally entangled with an ancillary one, denoted by $\mathcal{C}$. 
After that, the dynamics of $\mathcal{E}\cup \mathcal{S}$ is dictated by the same RUC for a time $t_2$. We are interested in the retrieval of this qubit in $\mathcal{E}$. Thus we study the mutual information
\be
I^{(2)}_{(c),\left[\mathcal{C},\mathcal{E}\right]}(t_2)={S}^{(2)}_\mathcal{C}(t_1+t_2)+{S}^{(2)}_\mathcal{E}(t_1+t_2)-{S}^{(2)}_{\mathcal{C}\cup \mathcal{E}}(t_1+t_2)\,,
\label{eq:mutual_renyi_c}
\ee
where the index $(c)$ here is used to distinguish this protocol from those in Fig.~\ref{fig:models_initial confuguration}. As usual, we average over the choice of the qudit $A$: this allows us, once again, to rely on the permutational symmetry in the four-replica space, and exploit the exact same techniques developed so far to efficiently simulate the dynamics (cf. Appendix~\ref{sec:details_c}). We report our numerical results in Figs.~\ref{fig:mutual_info_conserved_later_times} and \ref{fig:mutual_info_conserved_later_times_II}, which we now discuss.
	
First, Fig.~\ref{fig:mutual_info_conserved_later_times}$(a)$ displays the mutual information $	I^{(2)}_{(c),\left[\mathcal{C},\mathcal{E}\right]}(t_2)$ for increasing values of $t_1$. The plot corresponds to $N=40$ and initial charge $Q=N-1=39$, while $\lambda_1=1$, $\lambda_2=2$. For this choice of the parameters, the Page time is $t_p\sim 25$ [cf. Fig.~\ref{fig:mutual_info_conserved}$(a)$], so that $t_1<t_p$ for the data reported in Fig.~\ref{fig:mutual_info_conserved_later_times}$(a)$ . In this case, we see that  $I_{(c),[\mathcal{C},\mathcal{E}]}(t_2)$ saturates faster as time increases, which is what we expect. Indeed, for $t_1$ smaller than the Page time $t_{p}$, the RUC increases the entanglement between $\mathcal{S}$ and $\mathcal{E}$, so a retriever accessing $\mathcal{E}$ at time $t=t_1$ has more control over the configuration of the ``black hole'' when the extra qubit is injected.
	
At $t_1\sim t_{p}$, $\mathcal{S}$ and $\mathcal{E}$ will be maximally entangled  within a given charge sector. Thus, the retriever should be able to faithfully recollect information on the injected qubit after the scrambling time $t_s$. However, since the charge is conserved, the portion of the Hilbert space that can be explored during the dynamics is smaller that $2^N$. For this reason, we expect $t_s\propto \log S$, where $S=\ln D_{BH}(t_p)$ and $D_{BH}(t_p)$ is the effective dimension defined in Eq.~\eqref{eq:effective_HS}. Unfortunately, we can not reach large enough system sizes to test this statement quantitatively.

Next, we report in Fig.~\ref{fig:mutual_info_conserved_later_times}$(b)$ the mutual information $I^{(2)}_{(c),\left[\mathcal{C},\mathcal{E}\right]}(t_2)$ for $t_1>t_p$ and fixed initial charge $Q$. The plot shows that as $t_1$ increases the mutual information saturates more slowly, which is due to the fact that the entanglement between $\mathcal{S}$ and $\mathcal{E}$ decreases for $t_1>t_p$. In this respect, it is particularly simple to understand the limit $t_1\to\infty$: in this case the configuration of $\mathcal{S}$ at time $t=t_1$ will be extremely close to the vacuum, and there will be essentially no scrambling of information in $\mathcal{S}$, leading to an extremely slow saturation of $I^{(2)}_{(c),\left[\mathcal{C},\mathcal{E}\right]}(t_2)$ .

From Fig.~\ref{fig:mutual_info_conserved_later_times}$(b)$ we can also extract the dependence of the scrambling time for information injected at time $t_1$ on the system R\'enyi-$2$ entropy at time $t_1$, namely $S^{(2)}_{N}(t_1)$ . Here the scrambling time $t_s(t_1)$ is defined as the value of $t_2$ at which the mutual information reaches the value $2-\delta$, where $\delta$ is some small positive number. This is reported in Fig.~\ref{fig:mutual_info_conserved_later_times_II}$(a)$, where we chose $\delta=0.2$. From the plot it is clear that  $t_s(t_1)$ is a monotonically decreasing function of $S^{(2)}_{N}(t_1)$ for $t_1>t_p$, as we already discussed above.

Finally,  Fig.~\ref{fig:mutual_info_conserved_later_times_II}$(b)$ shows $I^{(2)}_{(c),\left[\mathcal{C},\mathcal{E}\right]}(t_2)$ for different values of the initial charge $Q$, for fixed $t_1\sim t_p(Q=39)$ (the Page time depends on the initial charge). In this case, we see that $I^{(2)}_{(c),\left[\mathcal{C},\mathcal{E}\right]}(t_2)$ is decreasing with $Q$, which is what we expect: if the initial charge is small, then the corresponding Page time is short. So, for $Q<39$ and a given time $t_1>t_p(Q=39)$, the entanglement between $\mathcal{S}$ and $\mathcal{E}$ will be small, leaving the retriever with little control over the configuration of $\mathcal{S}$ when the extra qubit is injected.

	\section{Conclusions}
	\label{sec:conclusions}
	
	In this work, we have considered the dynamics of a quantum many-body qudit system coupled to an external environment, where the time evolution is driven by the continuous limit of certain $2$-local random unitary circuits. We have shown that the  growth of the second R\'enyi entropy displays two different time scales that are related to the internal information scrambling and the interaction with the environment. Furthermore, we have characterized the qualitative differences that emerge choosing the unitaries to be Haar-distributed or with a conserved $U(1)$ charge. In the latter case, we have shown that the entanglement displays a Page-like behavior in time, where it begins to decrease in the middle stage of the ``evaporation''. Finally, we have shown that our model provides a microscopic realization of the Hayden-Preskill protocol for information retrieval, studying quantitatively the time evolution of the mutual information between different subsystems. The conserved $U(1)$ charge provides a tunable effective Hilbert space size, and allow us to study the charge dependence of scrambling dynamics.
	
	The RUC considered in this work can be  enriched in a number of ways. For instance, we have always considered the limit where the environment has an infinite number of qudits, that are non-interacting with one another. One could wonder whether the qualitative features described in this  work are modified by considering an environment with a finite number of qudits, possibly with a non-trivial internal dynamics.
	
	Next, it  would be extremely interesting to consider the growth of local operators~\cite{shenker_black_2014, shenker_multiple_2014,Kitaev_talk, roberts_localized_2015,maldacena2016bound,roberts_operator_2018} in our setting. While the effect of decoherence on the latter has been already considered in the literature~\cite{Yoshida2019Disentangling}, our model provides an ideal playground where numerical and analytic results can be derived for large values of $N$, and the implications of conservation laws explored in  detail. We plan to go back to these questions in future investigations.
	
	Finally, when compared to holographic duality, our model gives us a toy model for the boundary dynamics. It would be interesting to use a tensor network approach to describe bulk degrees of freedom, and study the entanglement wedge structure.

	%%%%%%%%%%%%%%%%%%%
	%%%%%%%%%%%%%%%%%%%
	\section*{Acknowledgments} 
	%%%%%%%%%%%%%%%%%%%
	%%%%%%%%%%%%%%%%%%%
	
We are very grateful to Ignacio Cirac, Vedika Khemani, Norbert Schuch,  for early collaboration on this project and valuable comments on this manuscript. We thank Masamichi Miyaji, Zhenbin Yang, Shunyu Yao, Shangnan Zhou for helpful discussions. LP acknowledges support from the Alexander von Humboldt foundation. XLQ is supported by the National Science Foundation under grant No. 1720504, the Simons Foundation, and in part by the Department of Energy under grant No. DE-SC0019380. 
	
	\appendix
	
	\section{Derivation of the  Lindbladian for the R\'enyi entropies}
	\label{sec:lindbladian}
	
	We wish to write down an evolution equation for the state $\ket{\rho_{\mathcal{S}}(t)\otimes \rho_{\mathcal{S}}(t)}\rangle$. To this end, we start with the discrete version of the quantum circuit introduced in Sec.~\ref{sec:model}. Choosing a time $t_j$ fixed, we focus on an individual realization of the circuit. This defines a global unitary transformation on $\mathcal{S}\cup \mathcal{E}$ which we denote by $U(t_j)$. Then, we have
	\be
	\ket{\rho_{\mathcal{S}}(t_j+\Delta t)\otimes  \rho_{\mathcal{S}}(t_j+\Delta t)}\rangle=\mathbb{1}_{\mathcal{H}_{\mathcal{S}}}\otimes  {\rm tr}_{\mathcal{E}}\left\{U(t_j+\Delta t) \rho U^{\dagger}(t_j+\Delta t) \right\}\otimes\mathbb{1}_{\mathcal{H}_{\mathcal S}}\otimes  {\rm tr}_{\mathcal{E}}\left\{U(t_j+\Delta t) \rho U^{\dagger}(t_j+\Delta t) \right\}\ket{I^+}_{1,\ldots,N}\,.
	\label{label:start}
	\ee
	The operator $U(t_j+\Delta t) $ is obtained from $U(t_j) $ by applying a suitable unitary operator. In particular, according to the evolution described in Sec.~\ref{sec:model}, we have three possibilities:
	\begin{itemize}
		\item with probability $1-p_1-p_2$ no unitary is applied at time $t_j$, so that $U(t_j+\Delta t)= U(t_j)$;
		\item with probability $p_1$ a unitary between $j$ and  $k$ is applied, so that  $U(t_j+\Delta t)=U_{j,k}U(t)$
		\item with probability $p_2$ a swap exchanges one qudit in $\mathcal{S}$ and one qudit in $\mathcal{E}.$
	\end{itemize}
	We can now take the average over all possible realizations. We note that the average can be taken independently at each time step, so that, due to the above considerations, the r. h. s. of Eq.~\eqref{label:start} splits into the sum of three contributions
	\be
	\mathbb{E}\left[\ket{\rho_{\mathcal{S}}(t_j+\Delta t)\otimes  \rho_{\mathcal{S}}(t_j+\Delta t)}\rangle\right]=(1-p_1-p_2)C_1+p_1 C_2+ p_2C_3\,.
	\ee
	The first, corresponding to no unitary applied, is trivial
	\be
	C_1=\mathbb{E}\left[\ket{\rho_{\mathcal{S}}(t_j)\otimes  \rho_{\mathcal{S}}(t_j)}\rangle\right]\,.
	\ee
	Next, $C_2$ can be easily determined, since the action of $U_{j,k}$, for $j,k\in\mathcal{S}$, commutes with tracing over $\mathcal{E}$. We obtain
	\be
	C_2=\frac{2}{N(N-1)}\sum_{j<k}\mathbb{E}\left[U^{\ast}_{j,k}\otimes U_{j,k}\otimes U^{\ast}_{j,k}\otimes U_{j,k}\right]\mathbb{E}\left[\ket{\rho_{\mathcal{S}}(t_j)\otimes  \rho_{\mathcal{S}}(t_j)}\rangle\right]\,.
	\ee
	The term $C_3$ is more complicated, because it couples the system $\mathcal{S}$ and the environment $\mathcal{E}$. However, it can be computed explicitly in the limit $M\to\infty$. Indeed, let us denote by $j$ and $k$ the qudits in $\mathcal{S}$ and $\mathcal{E}$ respectively that are swapped at time $t_j$. Assuming $M\gg N t_j/\Delta t$, we have a negligible probability that qudit $k$ in the environment has  interacted before with $\mathcal{S}$. Hence, we can assume $k$ to be in its initial configuration $\ket{0}_k$, and hence having no entanglement with the  rest of  the qudits in  $\mathcal{E}$. Under this assumption (which becomes exact in the limit $M\to \infty$), it is straightforward to compute
	\be
	C_3=\frac{1}{N}\sum_{j=1}^N\ket{0,0,0,0}_j\bra{I^+}_j \mathbb{E}\left[\ket{\rho_{\mathcal{S}}(t_j)\otimes  \rho_{\mathcal{S}}(t_j)}\rangle\right]\,,
	\ee 
	where $\ket{I^+}_j$ was introduced in \eqref{eq:def I+}.
	Putting all together and scaling the probabilities $p_1$ and $p_2$ with $\Delta t$ and $N$ as defined in \eqref{eq:p_1} and \eqref{eq:p_2} results in
	\begin{equation}
	\mathbb{E}\left[\ket{\rho_{\mathcal{S}}(t_j+\Delta t)\otimes  \rho_{\mathcal{S}}(t_j+\Delta t)}\rangle\right]=(1-\Delta t \mathcal{L})\mathbb{E}\left[\ket{\rho_{\mathcal{S}}(t_j)\otimes  \rho_{\mathcal{S}}(t_j)}\rangle\right]
	\end{equation}
	with the final Lindbladian \eqref{eq:lindbladian_final} as $\mathcal{L}$. In the limit $\Delta t \to 0$ we recover the differential equation \eqref{eq:diff_eq}.

	\section{Derivation of the system of differential equations for the purity in the Haar-scrambled case}
	\label{sec:purity_diff_eq}
	In the maximally chaotic case, we do not need to evaluate directly \eqref{eq:diff_eq} to obtain $\ket{\rho_{\mathcal S}(t)\otimes\rho_{\mathcal{S}}(t)}\rangle$. Instead, we can derive the system \eqref{eq:purtiy_equation} of $N+1$ coupled differential equations for the purities $\mathcal{P}_n = \langle\braket{W_n|\rho_{\mathcal{S}}\otimes\rho_{\mathcal S}}\rangle$ (see \eqref{eq:def_purity}) for subsystems of size $n$.
	
	To this end, we insert the Lindbladian \eqref{eq:lindbladian_final} into the equation \eqref{eq:def_purity} defining the purity,
	\begin{equation}
	\frac{d \mathcal{P}_n(t)}{dt} = \langle\langle W_n| (-\mathcal{L}) | \rho_{\mathcal S}(t)\otimes\rho_{\mathcal S}(t)\rangle\rangle. \end{equation}
	Next, the action of $\langle\langle W_n |$ from \eqref{eq:vector_form} onto the Lindbladian $\mathcal L$ from \eqref{eq:lindbladian_final} with $\mathcal{U}_{i,j}$ from \eqref{eq:four_u_haar} can be computed.
	Using the identities
	\begin{equation}
	\braket{I^\pm_j|I^\pm_k} = d^2\delta_{jk},\ \braket{I^\pm_j|I^\mp_k} = d\delta_{jk},\ \text{and}\ \braket{I^\pm_j|0,0,0,0_k} = \delta_{jk}\,,
	\end{equation}
	and keeping in mind that for $\langle\bra{W_n} = \langle\bra{W_K}$ only the size $|K|=n$ of the region matters, this results in
	\begin{align}
	&\langle\langle W_n | (-\mathcal{L})\nonumber \\
	& = -\frac{2\lambda_1}{N-1} \Bigg( \frac{N(N-1)}{2}\langle\langle W_n| - \frac{n(n-1)}{2} \frac{1}{d^4-1}\left[d^2\langle\bra{W_{n-2}} + d^4\langle\bra{W_{n}} - \frac{1}{d^2}\left(d^2\langle\bra{W_{n}} + d^4\langle\bra{W_{n-2}}\right) \right] \nonumber \\
	&\hspace{5em} - \frac{(N-n)(N-n-1)}{2}\frac{1}{d^4-1}\left[d^4\langle\bra{W_{n}} + d^2\langle\bra{W_{n+2}} - \frac{1}{d^2}\left(d^4\langle\bra{W_{n+2}} + d^2\langle\bra{W_{n}}\right)\right]\nonumber \\
	&\hspace{5em} - n(N-n)\frac{1}{d^4-1}\left[d^3 \langle\bra{W_{n-1}} + d^3\langle\bra{W_{n+1}} - \frac{1}{d^2}\left(d^3\langle\bra{W_{n+1}} + d^3\langle\bra{W_{n-1}}\right)\right] \Bigg)\nonumber \\
	&\quad -\frac{\lambda_2}{N} \left( N \langle\bra{W_n} - n\langle\bra{W_{n-1}} - (N-n) \langle\bra{W_{n}}  \right) \nonumber\\
	&= -\frac{2\lambda_1}{N-1}\left(n(N-n)\langle\bra{W_n} - \frac{n(N-n)d}{d^2+1}\left[\langle\bra{W_{n-1}} + \langle\bra{W_{n+1}}\right]\right) \nonumber\\
	&\quad -\frac{\lambda_2n}{N}\left(\langle\bra{W_{n}} - \langle\bra{W_{n-1}}\right)
	\end{align}
	by considering separately the three sets of terms in the sum $\sum_{1\le j<k\le N}$ where the $j$'th and $k$'th site of $\langle\langle W_n|$ consist of $\langle I^\pm|_j,\langle I^\pm|_k$ with the signs $+,+$ or $-,-$ or opposite, respectively.
	The differential equation \eqref{eq:purtiy_equation} for the purities $\mathcal{P}_n(t)$ then easily follows.
	
	\section{Derivation of the relevant formulas in the bosonic formalism}
	\label{sec:bosonic_modes}
	
	In this section we discuss in more detail the formalism introduced in Sec.~\ref{sec:entropy_u1}, and derive a set of formulas that are needed for numerical implementations. We start by showing how to write operators in terms of the bosonic $a$-operators. First, we notice that one simply has
	\be
	\sum_{j=1}^N\left(\ket{x}\bra{y}\right)_j=a^{\dagger}_x a_y\,,
	\ee
	as can be explicitly checked by comparing the action of the two sides on any state. From this, is follows
	\bea
	\sum_{j<k}^N\left[\left(\ket{x}\bra{z}\right)_j\otimes \left(\ket{y}\bra{t}\right)_k+\left(\ket{y}\bra{t}\right)_j\otimes \left(\ket{x}\bra{z}\right)_k\right]&=&\left(\sum_{j=1}^N\left(\ket{x}\bra{z}\right)_j\right) \left(\sum_{j=1}^N\left(\ket{y}\bra{t}\right)_j\right) -\delta_{y,z}\sum_{j=1}^N\ket{x}\bra{t}\nonumber\\
	&=&a^{\dagger}_x a_z a^{\dagger}_y a_t-\delta_{y,z} a^{\dagger}_x a_t\nonumber\\
	&=&a^{\dagger}_x a^{\dagger}_y a_z a_t\,.
	\label{eq:four_body_id}
	\eea
	for $x,y,z,t=\mathbf{0},\mathbf{1}, A, B, C, D$. One can now prove a general formula, which can be directly applied for implementing the effective Hamiltonians appearing in the main text. Let us consider
	\be
	\sum_{j<k}\sum_{x,y,z,t}\left(\Gamma_{x,y}\ket{x}_j\otimes \ket{y}_k\right) \left(\Lambda_{z,t}\bra{z}_j\otimes \bra{t}_k\right)=:(\ast)\,,
	\ee
	where $\Gamma_{x,y}=\Gamma_{y,x}$ and $\Lambda_{z,t}=\Lambda_{t,z}$ are symmetric matrices. We can rewrite
	\bea
	(\ast)&=&\sum_{j<k}\sum_{x,y,z,t}\Gamma_{x,y}\Lambda_{z,t} \left(\ket{x}\bra{z}\right)_j \otimes  \left(\ket{y} \bra{t}\right)_k\,,\nonumber\\
	&=&\frac{1}{2}\sum_{j<k}\sum_{x,y,z,t}\Gamma_{x,y}\Lambda_{z,t}\left[ \left(\ket{x}\bra{z}\right)_j \otimes  \left(\ket{y} \bra{t}\right)_k+ \left(\ket{y}\bra{t}\right)_j \otimes  \left(\ket{x} \bra{z}\right)_k\right]\nonumber\\
	&+&\frac{1}{2}\sum_{j<k}\sum_{x,y,z,t}\Gamma_{x,y}\Lambda_{z,t}\left[ \left(\ket{x}\bra{z}\right)_j \otimes  \left(\ket{y} \bra{t}\right)_k-\left(\ket{y}\bra{t}\right)_j \otimes  \left(\ket{x} \bra{z}\right)_k\right]\,.
	\eea
	In the second term, the parenthesis that multiplies $\Gamma_{x,y}\Lambda_{z,t}$ is antisymmetric under simultaneous exchange $x\leftrightarrow y$, $z\leftrightarrow t$. Since $\Gamma_{x,y}\Lambda_{z,t}$ is instead symmetric, the sum is zero. Accordingly, we have
	\bea
	(\ast)&=&\frac{1}{2}\sum_{j<k}\sum_{x,y,z,t}\Gamma_{x,y}\Lambda_{z,t}\left[ \left(\ket{x}\bra{z}\right)_j \otimes  \left(\ket{y} \bra{t}\right)_k+ \left(\ket{y}\bra{t}\right)_j \otimes  \left(\ket{x} \bra{z}\right)_k\right]=\frac{1}{2}\sum_{x,y,z,t}\Gamma_{x,y}\Lambda_{z,t}\left[a^{\dagger}_x a^{\dagger}_y a_z  a_t\right]\,.
	\eea
	where we used Eq.~\eqref{eq:four_body_id}.
	
	Finally, we show how to write symmetrized states in terms of bosonic $a$-operators. For this we consider a general state described by coefficients $c_{i,z}$, $i\in\{1,\ldots,N\}, z\in\{\mathbf0,\mathbf1,A,B,C,D\}$, which we symmetrize:
	\begin{align}
	&\frac{1}{N!}\sum_{\pi\in S_N} \pi \bigotimes_{i=1}^N\left(c_{i,\mathbf0}\ket{\mathbf  0}_i + c_{i,\mathbf1}\ket{\mathbf 1}_i + c_{i,A}\ket{A}_i+\cdots\right) \pi^{-1}\nonumber\\
	&= \frac{1}{N!}\sum_{\pi\in S_N} \pi \sum_{I_{\mathbf0}\cup I_{\mathbf1}\cup I_A \cup\cdots = \{1\ldots N\}} c_{I_{\mathbf{0}},\mathbf0}\ket{\mathbf 0}^{\otimes I_{\mathbf0}} c_{I_{\mathbf1},\mathbf1}\ket{\mathbf 1}^{\otimes I_{\mathbf1}} c_{I_A,A}\ket{A}^{\otimes I_A} \cdots \pi^{-1}\nonumber\\
	&= \frac{1}{N!}\sum_{I_{\mathbf0}\cup I_{\mathbf1}\cup I_A \cup\cdots = \{1\ldots N\}} \sqrt{N!\#I_{\mathbf0}!\#I_{\mathbf1}!\#I_A!\cdots} c_{I_{\mathbf0},\mathbf0}c_{I_{\mathbf1},\mathbf1}c_{I_A,A}\cdots\ket{\#I_{\mathbf0},\#I_{\mathbf1},\#I_A,\ldots}\nonumber \\
	&=  \frac{\sqrt{N!}}{N!} \sum_{I_{\mathbf0}\cup I_{\mathbf1}\cup I_A \cup\cdots = \{1\ldots N\}} c_{I_{\mathbf0},\mathbf0}(a_{\mathbf0}^\dag)^{\#I_{\mathbf0}} c_{I_{\mathbf1},\mathbf1}(a_{\mathbf1}^\dag)^{\#I_{\mathbf1}}c_{I_A,A}(a_A^\dag)^{\#I_A}\cdots \ket{\Omega}\nonumber\\
	&=\frac{1}{\sqrt{N!}}\prod_{i=1}^N (c_{i,\mathbf0}a_{\mathbf0}^\dag + c_{i,\mathbf1}a_{\mathbf1}^\dag + c_{i,A}a_A^\dag+\cdots)\ket{\Omega}    
	\label{eq:general_formula}
	\end{align}
	where $c_{I,z} = \prod_{i\in I} c_{i,z}$.
	
	From the above general formulas, it is now straightforward to rewrite the Lindbladian~\eqref{eq:lindbladian_final}, with the choice~\eqref{eq:four_u_conserved}, in terms of bosonic operators, together with the states relevant for our computations. In particular, we derived
	\bea
	\sum_{j<k}\mathcal{U}_{j,k}&=\frac{1}{2}\left(\sum_{\alpha=\mathbf0,\mathbf1,A,B,C,D} a_\alpha^ \dagger   a_\alpha^ \dagger   a_\alpha 
	a_\alpha \right)
	+\sum_{\alpha=A,B,C,D} \left(a_{\mathbf0}^ \dagger   a_\alpha^ \dagger   a_{\mathbf0}  
	a_\alpha +a_{\mathbf1}^ \dagger   a_\alpha^ \dagger   a_{\mathbf1} a_\alpha \right) \nonumber\\
	&+ \frac{1}{3} \left( a_{\mathbf0}^ \dagger   a_{\mathbf1}^ \dagger   a_A    a_B +  a_A^ \dagger   a_B^ \dagger   a_{\mathbf0}  
	a_{\mathbf1} +  a_{\mathbf0}^ \dagger   a_{\mathbf1}^ \dagger   a_C    a_D + a_C^ \dagger   a_D^ \dagger   a_{\mathbf0}  
	a_{\mathbf1} \right)\nonumber\\
	&+\frac{1}{3}\left(2 a_{\mathbf0}^ \dagger   a_{\mathbf1}^ \dagger   a_{\mathbf0}    a_{\mathbf1} + 2  a_A^ \dagger   a_B^ \dagger   a_A  
	a_B  +2 a_C^ \dagger   a_D^ \dagger   a_C    a_D  - a_C^ \dagger   a_D^ \dagger   a_A    a_B -  a_A^ \dagger   a_B^ \dagger   a_C  
	a_D \right)\,,
	\eea
	and
	\be
	\sum_{j=1}^N\ket{0,0,0,0}_j \bra{I^+}_j=\sum_{j=1}^N\ket{\mathbf{0}}_j(\bra{\mathbf{0}}_j+\bra{\mathbf{1}}_j+\bra{A}_j+\bra{B}_j) = a_{\mathbf0}^\dag(a_{\mathbf0}+a_{\mathbf{1}}+a_A+a_B)\,.
	\ee
	Furthermore, it follows from Eq.~\eqref{eq:general_formula} that Eq.~\eqref{eq:vector_form} can be  rewritten in terms  of bosonic modes as
	\bea
	\ket{W_K}\rangle &=& \frac{1}{\sqrt{N!}}(a_\mathbf0^\dag + a_\mathbf1^\dag + a_A^\dag + a_B^\dag)^{N-k}(a_\mathbf0^\dag + a_\mathbf1^\dag + a_C^\dag + a_D^\dag)^{k}\ket{\Omega}\,,
	\label{eq:w_vector_bosonic}
	\eea
	where $k=|K|$. Note that Eq.~\eqref{eq:w_vector_bosonic} actually corresponds to symmetrizing over all possible sets $K$ of $k$ elements. This is correct, since we are interested in the overlap~\eqref{eq:def_purity}, and the state $\ket{\rho_{\mathcal{S}}(t)\otimes \rho_{\mathcal{S}}(t)}\rangle$ is invariant under arbitrary permutations.

	Finally, let us consider the initial state~\eqref{eq:initial_state}. It is immediate to see that this corresponds to the state
	\be
	\ket{\rho_\mathcal{S}(0)\otimes\rho_\mathcal{S}(0)} = \ket{n_{\mathbf0}=0, n_\mathbf1=N, n_A=0, n_B=0,n_C=0, n_D=0}\,.
	\ee
	Similarly, averaging the initial state~\eqref{eq:initial_state_halfhalf} over all the possible permutations of qubits, we obtain the initial state
	\begin{equation}
	\ket{\rho_\mathcal{S}(0)\otimes\rho_\mathcal{S}(0)} = \frac{(N/2)!}{\sqrt{N!}} \ket{n_\mathbf0 = N/2, n_\mathbf1=N/2, n_A=0, n_B=0, n_C=0, n_D=0}.
	\end{equation}

	\section{Details on the computation of  the mutual information}
	\label{sec:details_mutual_info}
	
	In this section we provide all the necessary details to obtain the results on information retrieval presented in Sec.~\ref{sec:info_retrieval}. For ease of presentation and numerical efficiency, we restrict to qubits.
	
	\subsection{Scenario $(a)$}
	Let us begin with scenario (a), in which the black hole is in an initial product state except for one qubit $A$ [cf. Fig.~\ref{fig:models_initial confuguration}$(a)$].
	Rather than the initial product state \eqref{eq:initial_state} for the system $\mathcal{S}$, the initial state is now an entangled state
	\begin{equation}
	\ket{\Psi_0^{\mathcal S \cup \mathcal C}} = \frac{1}{\sqrt{2}}\sum_{s=0,1} \bigotimes_{j=1}^{N-1}\ket{1}_j\otimes\ket{s}_A\otimes\ket{s}_{\mathcal C}\,,
	\label{eq:double_product}
	\end{equation}
	with the two-replica Jamiolkowski representation
	\begin{equation}
	\ket{\rho_{\mathcal{S}\cup\mathcal{C}}(0)\otimes\rho_{\mathcal{S}\cup\mathcal{C}}(0)}\rangle = \frac{1}{4}\sum_{s\in\{0,1\}^4}\bigotimes_{j=1}^{N-1} \ket{\mathbf 1}_j\otimes\ket{s}_A\otimes\ket{s}_\mathcal{C}\,.
	\end{equation}
	After time evolution, we may extract the purities necessary for the mutual information \eqref{eq:renyi_bob_doesnt_know_II} similarly to \eqref{eq:def_purity}, but the vector $\ket{W}\rangle$ is now defined on systems $\mathcal{S}$ and $\mathcal{C}$. In particular, for the various Rényi entropies needed, we have
	\bea
	\ket{W_\mathcal{C}}\rangle &=& \bigotimes_{j\in\mathcal{S}} \ket{I^+}_j \otimes \ket{I^-}_{\mathcal{C}}\,\\
	\ket{W_{\mathcal{S}\cup\mathcal{C}}}\rangle &=&\bigotimes_{j\in\mathcal{S}} \ket{I^{-}}_j \otimes\ket{I^-}_{\mathcal{C}}\,,\\
	\ket{W_{\mathcal{S}}}\rangle &=& \bigotimes_{j\in\mathcal{S}}\ket{I^-}_j \otimes\ket{I^+}_{\mathcal{C}}\,. \label{eq:w_ancilla}
	\eea
	In order to perform the calculation of the purities
	\begin{equation}
	\mathcal{P}_X(t) = \langle\langle W_X | e^{-\mathcal{L}t} | \rho_{\mathcal{S}\cup\mathcal{C}}(0)\otimes\rho_{\mathcal{S}\cup\mathcal{C}}(0) \rangle\rangle,
	\end{equation}
	we symmetrize over system $\mathcal{S}$, including the location choice of qubit $A$ entangled to $\mathcal{C}$. Due to the projection onto $\bra{I^\pm}_\mathcal{C}$, the sum over $s$ may be restricted to $s\in\{\mathbf 0, \mathbf 1, A, B, C, D\}$.
	
	In the maximally chaotic case, we can derive and use the differential equation~\eqref{eq:purtiy_equation} as in section~\ref{sec:entropy_no_cons}, with initial conditions
	\bea
	\mathcal{P}_{K\cup\mathcal C} = \frac{k}{N} + \frac{1}{2}\frac{N-k}{N}\,,\\
	\mathcal{P}_{K} = \frac{1}{2}\frac{k}{N} + \frac{N-k}{N}\,,
	\eea
	where $K\subset\mathcal{S}$, $k=|K|$ as usual.
	
For the case with conservation laws, the symmetrization allows us to express all the states in the four replica space in the bosonic formalism as in section~\ref{sec:entropy_u1}, within which we can numerically compute the purities. For this, one needs to write down explicitly an expression for $\ket{\rho_{\mathcal{S}\cup\mathcal{C}}(0)\otimes\rho_{\mathcal{S}\cup\mathcal{C}}(0)} \rangle $. To this end, let us generalize the case considered in Eq.~\eqref{eq:double_product}, by considering instead the case corresponding to the initial state~\eqref{eq:initial_state_n_dep}, where we sum over all the possible permutations of qubits. Then, following the technical derivations in the previous section, it is possible to derive
		\begin{equation}
		\ket{\rho_{\mathcal{S}\cup\mathcal{C}}(0)\otimes\rho_{\mathcal{S}\cup\mathcal{C}}(0)} \rangle = \frac{1}{4}\sum_{s\in\{\mathbf 0, \mathbf 1, A, B, C, D\}} \frac{1}{\sqrt{N!}} a_s^\dag (a_{\mathbf 1}^\dag)^{N-1-n}(a_{\mathbf{0}}^\dag)^{n}\ket{\Omega}_{\mathcal{S}}\otimes\ket{s}_{\mathcal C}.
		\end{equation}
		Note that here we have $N-1$, and not $N$, appearing  in the second exponent, because one qubit is  maximally entangled with  $\mathcal{C}$, so only $N-1$ qubits in $\mathcal{S}$ are in a product state.

	\subsection{Scenario $(b)$}
	Now let us move to scenario (b), in which the black hole is maximally entangled to a retriever $B$, except for one qubit $A$, that is maximally entangled to $\mathcal C$ [cf. Fig.~\ref{fig:models_initial confuguration}$(b)$]. Here, the initial state is an entangled state which reads
	\begin{equation}
	\ket{\rho_{\mathcal{S}\cup\mathcal{C}\cup B}(0)\otimes\rho_{\mathcal{S}\cup\mathcal{C}\cup B}(0)}\rangle = \frac{1}{4^N}\bigotimes_{j=1}^{N-1}\sum_{s_j\in\{0,1\}^4} (\ket{s_j}_{\mathcal{S},j}\otimes\ket{s_j}_{B,j})\otimes\sum_{s\in\{0,1\}^4}\ket{s}_A\otimes\ket{s}_\mathcal{C}.
	\end{equation}
	The $\ket{W}\rangle$ vectors for the purities involved in the mutual information \eqref{eq:renyi_bob_knows} are as in \eqref{eq:w_ancilla} with an additional $\ket{I^+}_{B,j}$ for each $j\in B$, since $B$ is never within a region we compute the purity of. Therefore we may directly evaluate $\sum_{s_j\in\{0,1\}^N}\ket{s_j}_{\mathcal S,j}\braket{I^+|s_j}_{B,j} = \ket{\mathbf 0} + \ket{\mathbf 1} + \ket{A} + \ket{B}$ and use the simplified initial state 
	\begin{align}
	\ket{\rho_{\mathcal{S}\cup\mathcal{C}}(0)\otimes\rho_{\mathcal{S}\cup\mathcal{C}}(0)} \rangle &= \frac{1}{4^N} \sum_{s\in\{0,1\}^4} \bigotimes_{j=1}^{N-1} (\ket{\mathbf 0}_j + \ket{\mathbf{1}}_j + \ket{A}_j+\ket{B}_j) \otimes \ket{s}_A\otimes\ket{s}_{\mathcal C} \\
	&= \frac{1}{4^N} \sum_{s\in\{\mathbf 0, \mathbf 1, A, B, C,D\}}\frac{1}{\sqrt{N!}}a_s^\dag(a_\mathbf0^\dag + a_\mathbf1^\dag + a_A^\dag+a_B^\dag)^{N-1}\ket{\Omega}_{\mathcal{S}}\otimes\ket{s}_{\mathcal C}
	\end{align}
	after restriction of $s$ and symmetrization of $\mathcal{S}$ as above. For the evolution with charge conservation, this bosonic formalism is again the basis for our numerical calculations.
	
	Note, finally, that in the case of Haar-scrambled evolution, we can again use the differential equation~\eqref{eq:purtiy_equation}, where the initial conditions are now
	\bea
	\mathcal{P}_{K\cup \mathcal C} &=& \frac{k}{N}\frac{1}{2^{k-1}} + \frac{N-k}{N}\frac{1}{2^{k+1}}\,,\\
	\mathcal{P}_{K} &=& \frac{1}{2^k}\,.
	\eea

	\subsection{Scenario $(c)$}
	\label{sec:details_c}
	In order to implement the two-step protocol depicted in Fig.~\ref{fig:model_later_times}, it is crucial to remember that the interaction with the bath is Markovian. First, we simply evolve for time $t_1$ the initial pure (and symmetrized) state \eqref{eq:initial_state_n_dep} of $N-1$ qubits in a given charge sector. Then, we add a qubit $s$ maximally entangled to the ancilla. Symmetrizing its position, this amounts to the following change of basis vectors of the system state:
	\begin{equation}
	    \ket{n_\mathbf0,n_\mathbf1,\ldots,n_D}_{\mathcal S} \to \frac{1}{4} \sum_{s\in\{\mathbf0,\mathbf1,A,B,C,D\}} \sqrt{\frac{n_s+1}{N}}\ket{n_\mathbf0 + \delta_{s\mathbf0}, n_\mathbf1+\delta_{s\mathbf1},\ldots,n_D+\delta_{sD}}_\mathcal{S} \otimes \ket{s}_{\mathcal{C}}.
	\end{equation}
	The rest of the protocol is then analogous to scenario (a) for time $t_2$ and the initial mixed state above.

	\bibliography{./bibliography}

\end{document}